\documentclass[11pt]{article}
\usepackage{amssymb}
\usepackage[dvips]{epsfig}
\usepackage{times}
\usepackage{graphicx}
\newtheorem{Lemma}{Lemma}[section]

\newtheorem{Proposition}[Lemma]{Proposition}
\newtheorem{Corollary}[Lemma]{Corollary}
\newtheorem{Remark}[Lemma]{Remark}

\newtheorem{Example}[Lemma]{Example}

\newenvironment{Proof}%
 {\begin{trivlist} \item[]{\bf Proof. }}%
 {\hspace*{\fill}$\rule{.3\baselineskip}{.35\baselineskip}$\end{trivlist}}
 {\begin{trivlist}\item[]\textbf{Acknowledgments }}{\end{trivlist}}

\makeatletter \@addtoreset{equation}{section} \makeatother

\newfam\bifam
\font\tenbi=cmmib10 scaled \magstep1 \font\sevenbi=cmmib10 at 11pt
\font\fivebi=cmmib10 at 6pt \textfont\bifam = \tenbi
\scriptfont\bifam= \sevenbi \scriptscriptfont\bifam= \fivebi

\begin{document}

\title{\bf Translationally invariant nonlinear Schr\"{o}dinger lattices}

\author{Dmitry E. Pelinovsky \\
{\small Department of Mathematics, McMaster University, Hamilton,
Ontario, Canada, L8S 4K1} }

\date{\today}
\maketitle

\begin{abstract}
Persistence of stationary and traveling single-humped localized 
solutions in the spatial discretizations of the nonlinear 
Schr\"{o}dinger (NLS) equation is addressed. The discrete NLS 
equation with the most general cubic polynomial function is 
considered. Constraints on the nonlinear function are found from the 
condition that the second-order difference equation for stationary 
solutions can be reduced to the first-order difference map. The 
discrete NLS equation with such an exceptional nonlinear function is 
shown to have a conserved momentum but admits no standard 
Hamiltonian structure. It is proved that the reduction to the 
first-order difference map gives a sufficient condition for 
existence of translationally invariant single-humped stationary 
solutions and a necessary condition for existence of single-humped 
traveling solutions. Other constraints on the nonlinear function are 
found from the condition that the differential advance-delay 
equation for traveling solutions admits a reduction to an integrable 
normal form given by a third-order differential equation. This 
reduction also gives a necessary condition for existence of 
single-humped traveling solutions. The nonlinear function which 
admits both reductions defines a two-parameter family of discrete 
NLS equations which generalizes the integrable Ablowitz--Ladik 
lattice. 
\end{abstract}

\section{Introduction}

We address spatial discretizations of the nonlinear Schr\"{o}dinger
(NLS) equation in one dimension,
\begin{equation}
\label{NLS} i u_t + u_{xx} + 2 |u|^2 u = 0, \qquad x \in \mathbb{R}, 
\quad t \in \mathbb{R}
\end{equation}
which has a family of traveling wave solutions
\begin{equation}
\label{soliton} u = \sqrt{\omega} \; {\rm sech}(\sqrt{\omega} (x - 2
c t - s)) \; e^{ i c (x - c t) + i \omega t + i \theta},
\end{equation}
where $\omega \in \mathbb{R}_+$ and $(c,s,\theta) \in \mathbb{R}^3$
are free parameters of the solution family. The discrete counterpart
of the NLS equation takes the form:
\begin{equation}
\label{dNLS} i \dot{u}_n + \frac{u_{n+1}-2u_n + u_{n-1}}{h^2} + 
f(u_{n-1},u_n,u_{n+1}) = 0, \qquad n \in \mathbb{Z}, \quad t \in 
\mathbb{R},
\end{equation}
where $h^2 > 0$ is parameter, and $f : \mathbb{C}^3 \mapsto
\mathbb{C}$ is a non-analytic function with the properties:
\begin{itemize}
\item[P1] (symmetry) $f(v,u,w) = f(w,u,v)$ \item[P2] (continuity)
$f(u,u,u) = 2 |u|^2 u$ \item[P3] (gauge covariance)
$f(e^{i\alpha}v,e^{i\alpha} u,e^{i\alpha} w) = e^{i\alpha}
f(v,u,w)$  $\forall \alpha \in \mathbb{R}$ \item[P4]
(reversibility) $\overline{f(v,u,w)} = f(\bar{v},\bar{u},\bar{w})$
\end{itemize}
We refer to the model (\ref{dNLS}) as the {\em discrete NLS
equation} or simply {\em the NLS lattice}. The discrete NLS equation
(\ref{dNLS}) is a symplectic semi-discretization of the continuous 
NLS equation (\ref{NLS}), where the second partial derivative is 
replaced with the second-order central difference on the grid $x = 
nh$, $n \in \mathbb{Z}$ and the nonlinearity incorporates the 
effects of on-site and adjacent-site couplings \cite{symplecticNLS}. 
From another point of view, the discrete model (\ref{dNLS}) is 
derived in various branches of physics, most recently, in the 
context of Bose--Einstein condensates in periodic optical lattices 
\cite{Kevrekidis-review}.

The symmetry property P1 is required to ensure that the 
semi-discretization is symplectic. The continuity property P2 is 
needed if the discrete NLS equation (\ref{dNLS}) is to be matched to 
the continuous NLS equation (\ref{NLS}) in the limit $h \to 0$. The 
gauge covariance and reversibility properties P3--P4 originate from 
applications of the discrete and continuous NLS equations to the 
modelling of the envelope of modulated nonlinear dispersive waves in 
a non-dissipative system \cite{Kevrekidis-review}. We enforce two 
additional properties on the nonlinear function $f$:
\begin{itemize}
\item[P5] $f(v,u,w)$ is independent on $h$ \item[P6] $f(v,u,w)$ is
a homogeneous cubic polynomial in $(v,u,w)$
\end{itemize}
Neither analysis nor applications require properties P5--P6. 
However, we use them to limit the search of all possible exceptional 
NLS lattices to a finite-dimensional parameter space. Indeed, we 
find immediately that all properties P1--P6 are satisfied if and 
only if the nonlinear function $f(u_{n-1},u_n,u_{n-1})$ is 
represented by the ten-parameter family of cubic polynomials,
\begin{eqnarray}
\nonumber f & = & \alpha_1 |u_n|^2 u_n + \alpha_2 |u_n|^2 (u_{n+1} +
u_{n-1}) + \alpha_3 u_n^2 (\bar{u}_{n+1} + \bar{u}_{n-1}) + \alpha_4
(|u_{n+1}|^2 + |u_{n-1}|^2) u_n \\
\nonumber & \phantom{t} & + \alpha_5 (\bar{u}_{n+1} u_{n-1} + 
u_{n+1} \bar{u}_{n-1} ) u_n + \alpha_6 (u_{n+1}^2 + u_{n-1}^2) 
\bar{u}_n + \alpha_7 u_{n+1} u_{n-1} \bar{u}_n \\
\nonumber & \phantom{t} & + \alpha_8 (|u_{n+1}|^2 u_{n+1} + 
|u_{n-1}|^2 u_{n-1}) + \alpha_9 (u_{n+1}^2 \bar{u}_{n-1} + \bar{u}_{n+1} u_{n-1}^2) \\
& \phantom{t} & + \alpha_{10} (|u_{n+1}|^2 u_{n-1} + |u_{n-1}|^2 
u_{n+1}), \label{nonlinearity}
\end{eqnarray}
where the real-valued parameters $(\alpha_1,...,\alpha_{10})$
satisfy the continuity constraint:
\begin{equation}
\label{constraint} \alpha_1 + 2 \alpha_2 + 2 \alpha_3 + 2 \alpha_4 +
2 \alpha_5 + 2 \alpha_6 + \alpha_7 + 2 \alpha_8 + 2 \alpha_9 + 2
\alpha_{10} = 2.
\end{equation}
This family generalizes the cubic on-site lattice when 
\begin{equation}
\label{onsiteNLS} {\rm (dNLS)} \qquad f = 2 |u_n|^2 u_n
\end{equation}
and the integrable Ablowitz--Ladik (AL) lattice
\begin{equation}
\label{AL} {\rm (AL)} \qquad f = |u_n|^2 ( u_{n+1} + u_{n-1} ). 
\end{equation}
A more general nonlinear function of the form (\ref{nonlinearity}) 
was derived in the context of modeling of the Fermi-Pasta-Ulam (FPU)
problem (see Eqs. (11)-(12) in \cite{CKKS93}) with 
\begin{equation}
\label{FPU} \alpha_1 = \alpha_2 = \alpha_4 = \frac{1}{4}, \qquad 
\alpha_3 = \alpha_6 = \alpha_8 = \frac{1}{8},\qquad \alpha_5 = 
\alpha_7 = \alpha_9 = \alpha_{10} = 0.
\end{equation}
We shall consider {\em persistence} of traveling wave solutions
(\ref{soliton}) in the discrete NLS equation (\ref{dNLS}) for 
sufficiently small $h$ depending on the nonlinear function 
(\ref{nonlinearity}). Since both the translational and Gallileo 
invariances are broken, existence of a {\em stationary} solution for 
some $\omega \in I_1 \subset \mathbb{R}$,
\begin{equation}
\label{stationary} u_n(t) = \phi(h n) e^{i \omega t}, \qquad \phi :
\mathbb{Z} \mapsto \mathbb{C},
\end{equation}
does not guarantee existence of a {\em traveling} solution for some
$(\omega,c) \in I_2 \subset \mathbb{R}^2$,
\begin{equation}
\label{travelling} u_n(t) = \phi(hn - 2 ct) e^{i \omega t}, \qquad
\phi : \mathbb{R} \mapsto \mathbb{C},
\end{equation}
where $\omega$ is frequency and $c$ is velocity of traveling 
solutions. In both cases (\ref{stationary}) and (\ref{travelling}), 
we are only interested in existence of {\em single-humped localized} 
solutions $\phi(z)$, which correspond to the ${\rm sech}$-solutions 
(\ref{soliton}) in the limit $h \to 0$. In what follows, {\em 
stationary} and {\em traveling} solutions stand for single-humped 
localized solutions unless the opposite is explicitly specified. 
Direct substitutions of (\ref{stationary}) and (\ref{travelling}) to 
the discrete NLS equation (\ref{dNLS}) show that the sequence $\{ 
\phi_n \}_{n \in \mathbb{Z}}$ with $\phi_n = \phi(hn)$ satisfies the 
second-order difference equation,
\begin{equation}
\label{second-order-difference} \frac{\phi_{n+1} - 2 \phi_n +
\phi_{n-1}}{h^2} - \omega \phi_n + f(\phi_{n-1},\phi_n,\phi_{n+1})
= 0, \quad n \in \mathbb{Z},
\end{equation}
while the function $\phi(z)$ with $z = h n - 2 ct$ satisfies the 
differential advance-delay equation,
\begin{equation}
\label{advance-delay} 2 i c \phi'(z) = \frac{\phi(z+h) - 2 \phi(z)
+ \phi(z-h)}{h^2} - \omega \phi(z) +
f(\phi(z-h),\phi(z),\phi(z+h)), \quad z \in \mathbb{R}.
\end{equation}
Existence of stationary solutions of the second-order difference
equation (\ref{second-order-difference}) can be established for a
large class of nonlinearities (\ref{nonlinearity}) by using the
variational method \cite{pankov}. Two single-humped solutions exist 
in a general case: one solution is symmetric about a selected 
lattice node (say $n = 0$) and the other solution is symmetric about 
a midpoint between two adjacent nodes.

There are two obstacles for the stationary solution
(\ref{stationary}) to persist as the traveling solution
(\ref{travelling}). First, the solution of the stationary problem
(\ref{second-order-difference}) corresponds generally to a {\em
piecewise continuous} solution $\phi(z)$ of the traveling problem
(\ref{advance-delay}) on $z \in \mathbb{R}$ \cite{Hump}. Jump
discontinuities of $\phi(z)$ may occur in between two adjacent
nodes, e.g. between $z_n = h n$ and $z_{n+1} = (n+1)h$. Second, even 
if there exists a {\em continuous} solution $\phi(z)$ of the 
traveling problem (\ref{advance-delay}) with $c = 0$, this solution 
may not persist as a {\em continuously differentiable} solution 
$\phi(z)$ of the problem (\ref{advance-delay}) with $c \neq 0$. The 
stationary solution of the second-order difference equation 
(\ref{second-order-difference}) is said to be {\em translationally 
invariant} if the function $\phi_n = \phi(nh)$ on $n \in \mathbb{Z}$ 
can be extended to a one-parameter family of continuous solutions 
$\phi(z-s)$ on $z \in \mathbb{R}$ of the advance-delay equation 
(\ref{advance-delay}) with $c = 0$, where $s \in \mathbb{R}$ is an 
arbitrary translation parameter.

The persistence problem for traveling waves in discrete lattices was
recently considered in a number of publications. Kevrekidis
\cite{K03} suggested a method to define an exceptional nonlinear
function $f(u_{n-1},u_n,u_{n+1})$ from the condition that the
discrete NLS equation (\ref{dNLS}) preserves the {\em momentum}
invariant,
\begin{equation}
M = i \sum_{n\in\mathbb{Z}} \left( \bar{u}_{n+1} u_n - u_{n+1}
\bar{u}_n \right). \label{momentum}
\end{equation}
No characterization of the nonlinear functions (\ref{nonlinearity}) 
that preserve the momentum conservation (\ref{momentum}) was made in 
\cite{K03} except for the integrable AL lattice (\ref{AL}). In 
addition, it was not shown in \cite{K03} that the preservation of 
momentum (\ref{momentum}) guarantees the existence of a continuous 
function $\phi(z)$ in the solution (\ref{travelling}) with $c = 0$.

Oster, Johansson, and Eriksson \cite{OJE03} addressed the most
general {\em Hamiltonian} discrete NLS equation in the form,
\begin{equation}
\label{Hamiltonian} i \dot{u}_n = \frac{\partial H}{\partial
\bar{u}_n}, \qquad H = \sum_{n \in \mathbb{Z}} \left(
\frac{|u_{n+1}-u_n|^2}{h^2} - F(u_n,u_{n+1}) \right),
\end{equation}
where $F : \mathbb{C}^2 \mapsto \mathbb{R}$ is a symmetric potential 
function of quartic polynomials which satisfies the gauge invariance 
property: $F(e^{i\alpha}u,e^{i\alpha} w) = F(u,w)$ $\forall \alpha 
\in \mathbb{R}$. The dNLS lattice (\ref{onsiteNLS}) and the FPU 
lattice (\ref{FPU}) are particular examples of the Hamiltonian 
discrete NLS equation (\ref{Hamiltonian}). It was found in 
\cite{OJE03} that there exists a configuration in parameters of 
quartic polynomials $F(u_n,u_{n+1})$ which provides a reduced 
Peierls--Nabarro potential and enhanced mobility of localized modes. 
No analysis of the localized modes with enhanced mobility was 
developed in \cite{OJE03}.

Ablowitz and Musslimani \cite{AM03} outlines problems in the
numerical approximations for traveling wave solutions
(\ref{travelling}) with $c \neq 0$ in the context of earlier
contradictory publications (see also review in \cite{PR05}). An
asymptotic method was developed in \cite{AM03} to show that if the
solution exists for $c = 0$ it can be continued to non-zero values
of $c$ as perturbation series expansions in powers of $c$.

Pelinovsky and Rothos \cite{PR05} computed the normal form for 
bifurcations of traveling solutions (\ref{travelling}) near the 
special value of the parameter $c = 1/h$. The normal form is 
represented by the third-order ODE related to the third-order 
derivative NLS equation \cite{PR05}. Existence of embedded solitons 
in the third-order derivative NLS equation was considered in the 
past (see \cite{YA03,PY05} and reference therein) and the previous 
results suggested that the dNLS lattice (\ref{onsiteNLS}) did not 
support bifurcation of {\em single-humped} traveling solutions in a 
sharp contrast to the case of the AL lattice (\ref{AL}). ({\em 
Double-humped} traveling solutions in the normal form reduction were 
predicted for the dNLS lattice (\ref{onsiteNLS}) in \cite{PR05} and 
numerical approximations of the double-humped traveling solutions in 
the differential advance-delay equation (\ref{advance-delay}) were 
obtained in \cite{Champ}.)

We shall review the results and conjectures of the previous works 
\cite{AM03,K03,OJE03,PR05} in the context of the cubic nonlinear 
function (\ref{nonlinearity}). Our results are based on the recent 
works \cite{BOP05} and \cite{DKY05}, where similar ideas are 
developed for monotonic kinks of the discrete $\phi^4$ theory. See 
also \cite{OPB05,IP06} for other relevant results on monotonic kinks 
in discrete Klein--Gordon lattices. 

In Section 2, the class of {\em exceptional} nonlinear functions in 
the general family of cubic polynomials (\ref{nonlinearity}) is 
identified from the condition that the second-order difference 
equation (\ref{second-order-difference}) admits a reduction to the 
first-order difference equation. It is shown that the discrete NLS 
equation (\ref{dNLS}) with an exceptional nonlinearity preserves the 
momentum conservation (\ref{momentum}) but admits no Hamiltonian in 
the form (\ref{Hamiltonian}). 

In Section 3, it is proved that the first-order difference equation 
admits a one-parameter family of stationary solutions 
(\ref{stationary}) for any $\omega > 0$ and sufficiently small $h$, 
and these stationary solutions are {\em translationally invariant}. 
Unlike the case of monotonic kinks in \cite{BOP05}, the proof of 
existence of a translationally invariant single-humped localized 
solution is based on a construction of two sequences which depend 
continuously on the initial value. One sequence is monotonically 
decreasing to zero and the other sequence is monotonically 
increasing until the turning point and it decreases monotonically 
beyond the turning point. The main outcome of this analysis is a 
clear evidence that the reduction to the first-order difference 
equation gives {\em a sufficient condition} for existence of 
translationally invariant stationary solutions and {\em a necessary 
condition} for existence of traveling  solutions near $c = 0$ for 
$\omega > 0$. 

Section 4 tests the exceptional nonlinear functions by the reduction 
of the differential advance-delay equation (\ref{advance-delay}) to 
the normal form near the special values of parameters $\omega = 
(\pi-2)/h^2$ and $c = 1/h$. Reductions of the normal form to 
integrable Hirota \cite{H73} and Sasa-Satsuma \cite{SS91} equations 
and existence of single-humped solutions in the integrable normal 
form give {\em a necessary condition} for existence of traveling 
solutions in the differential advance-delay equation 
(\ref{advance-delay}) near the special values of parameters 
$(\omega,c)$. 

Section 5 presents the explicit form of the two-parameter 
exceptional nonlinear function (\ref{nonlinearity}) that passes 
through both tests and may contain families of traveling solutions. 
This function generalizes the AL lattice (\ref{AL}) which is known 
to admit exact traveling solutions between the two limits $c = 0$ 
and $c = 1/h$. Numerical test of persistence of traveling solutions 
in the differential advance-delay equation (\ref{advance-delay}) is
outlined as an open problem.

\section{Reductions to the first-order difference equation}

We shall consider the stationary solution (\ref{stationary}) of the 
discrete NLS equation (\ref{dNLS}), which satisfies the second-order 
difference equation (\ref{second-order-difference}). We obtain the 
constraints on the function $f(\phi_{n-1},\phi_n,\phi_{n+1})$ in the 
family of cubic polynomials (\ref{nonlinearity}) from the condition 
that the second-order difference equation 
(\ref{second-order-difference}) admits a conserved quantity
\begin{equation}
\label{first-order-difference} E = \frac{1}{h^2} 
|\phi_{n+1}-\phi_n|^2 - \frac{1}{2} \omega \left( \phi_n 
\bar{\phi}_{n+1} + \bar{\phi}_n \phi_{n+1} \right) +
g(\phi_n,\phi_{n+1}),
\end{equation}
where $g : \mathbb{C}^2 \mapsto \mathbb{R}$ is a non-analytic 
function and $E \in \mathbb{R}$ is constant in $n \in \mathbb{Z}$. 
Due to properties P1--P6 for the function 
$f(\phi_{n-1},\phi_n,\phi_{n+1})$, the nonlinear function 
$g(\phi_n,\phi_{n+1})$ satisfies the properties:
\begin{itemize}
\item[S1] (symmetry) $g(u,w) = g(w,u)$ 
\item[S2] (continuity)
$g(u,u) = |u|^4$ and $\frac{\partial g}{\partial \bar{u}}(u,u) = 
\frac{\partial g}{\partial \bar{w}}(u,u) = |u|^2 u$
\item[S3] (gauge variance) $g(e^{i\alpha} u,e^{i\alpha} w) = g(u,w)$  
$\forall \alpha \in \mathbb{R}$
\item[S4] (reversibility) $\overline{g(u,w)} = g(\bar{u},\bar{w})$
\item[S5] $g(u,w)$ is independent on $h$ 
\item[S6] $g(u,w)$ is a
homogeneous quartic polynomial in $(u,w)$
\end{itemize}
The most general polynomial $g(\phi_n,\phi_{n+1})$ that satisfies 
properties S1--S6 takes the form:
\begin{eqnarray}
\nonumber g & = & \gamma_1 (|\phi_n|^2 + |\phi_{n+1}|^2) ( 
\bar{\phi}_{n+1} \phi_n + \phi_{n+1}  \bar{\phi}_n)  + \gamma_2 
|\phi_n|^2 |\phi_{n+1}|^2 \\    \label{function-g} & \phantom{t} & + 
\gamma_3 (\phi_n^2 \bar{\phi}_{n+1}^2 + \bar{\phi}_n^2 \phi_{n+1}^2) 
+ \gamma_4 (|\phi_n|^4 + |\phi_{n+1}|^4),
\end{eqnarray}
where the real-valued parameters 
$(\gamma_1,\gamma_2,\gamma_3,\gamma_4)$ satisfy the continuity 
constraint: 
\begin{equation}
4 \gamma_1 + \gamma_2 + 2 \gamma_3 + 2 \gamma_4 = 1.
\end{equation}
In the limit $h \to 0$, the second-order difference equation 
(\ref{second-order-difference}) reduces to the second-order ODE:
\begin{equation}
\label{second-order-ODE} \phi'' - \omega \phi + 2 |\phi|^2 \phi = 0,
\end{equation}
while the first-order difference equation 
(\ref{first-order-difference}) transforms to the first integral of 
(\ref{second-order-ODE})
\begin{equation}
\label{first-order-ODE} E = |\phi'|^2 - \omega |\phi|^2 + |\phi|^4.
\end{equation}
The first-order ODE (\ref{first-order-ODE}) with $E = 0$ admits a 
unique localized solution $\phi(z) = \sqrt{\omega} \; {\rm 
sech}(\sqrt{\omega}(z - s))$ with $s \in \mathbb{R}$ and $\omega > 
0$, which agrees with the exact solution (\ref{soliton}) for $z = x$ 
and $c = 0$. Therefore, the first-order difference equation 
(\ref{first-order-difference}) is a symplectic discretization of the 
first-order ODE (\ref{first-order-ODE}). This precise methodology 
was used in \cite{DKY05} to construct {\em translationally invariant 
monotonic kinks} in the discrete $\phi^4$ equation. (A more general 
approach is reported in \cite{BOP05}.) We first specify on the 
correspondence between the difference equations 
(\ref{second-order-difference}) and (\ref{first-order-difference}) 
and then prove three elementary results on existence of the 
reduction (\ref{first-order-difference}) with the nonlinear function 
(\ref{function-g}) and its relation to conservation of $M$ in 
(\ref{momentum}) and $H$ in (\ref{Hamiltonian}).

\begin{Lemma}
Let the function $g(\phi_n,\phi_{n+1})$ with properties S1--S6 be 
related to the function $f(\phi_{n-1},\phi_n,\phi_{n+1})$ with 
properties P1--P6 by 
\begin{eqnarray*}
g(\phi_n,\phi_{n+1}) - g(\phi_{n-1},\phi_n) = \frac{1}{2} \left[ 
(\bar{\phi}_{n+1} - \bar{\phi}_{n-1} ) 
f(\phi_{n-1},\phi_n,\phi_{n+1}) + (\phi_{n+1} - \phi_{n-1} ) 
\overline{f(\phi_{n-1},\phi_n,\phi_{n+1})} \right].
\end{eqnarray*}
If the sequence $\{ \phi_n \}_{n \in \mathbb{Z}}$ is any solution of 
the second-order difference equation 
(\ref{second-order-difference}), it satisfies the first-order 
difference equation (\ref{first-order-difference}). If the sequence 
$\{ \phi_n \}_{n \in \mathbb{Z}}$ is a non-constant real-valued 
solution of the first-order difference equation 
(\ref{first-order-difference}), it also satisfies the second-order 
difference equation (\ref{second-order-difference}). 
\label{lemma-correspondence-equations}
\end{Lemma}

\begin{Proof}
By subtracting the first-order difference equation 
(\ref{first-order-difference}) with $n \equiv n$ from that with $n 
\equiv n - 1$, we have
\begin{eqnarray}
\nonumber \frac{1}{2} (\bar{\phi}_{n+1} - \bar{\phi}_{n-1}) \left(
\frac{\phi_{n+1} - 2 \phi_n + \phi_{n-1}}{h^2} - \omega \phi_n
\right) \\ \label{correspondence-integral} +   \frac{1}{2}
(\phi_{n+1} - \phi_{n-1}) \left( \frac{\bar{\phi}_{n+1} - 2
\bar{\phi}_n + \bar{\phi}_{n-1}}{h^2} - \omega \bar{\phi}_n
\right)+ g(\phi_n,\phi_{n+1}) - g(\phi_{n-1},\phi_n) = 0.
\end{eqnarray}
The first statement follows immediately from the relation between 
$f(\phi_{n-1},\phi_n,\phi_{n+1})$ and $g(\phi_n,\phi_{n+1})$. Let us 
now assume that the first-order equation 
(\ref{first-order-difference}) is satisfied and functions 
$f(\phi_{n-1},\phi_n,\phi_{n+1})$ and $g(\phi_n,\phi_{n+1})$ are 
related as above. Then, the non-constant sequence $\{ \phi_n \}_{n 
\in \mathbb{Z}}$ satisfies the second-order equation,
$$
\frac{\phi_{n+1} - 2 \phi_n + \phi_{n-1}}{h^2} - \omega \phi_n + 
f(\phi_{n-1},\phi_n,\phi_{n+1}) = i (\phi_{n+1}-\phi_{n-1}) 
h(\phi_{n-1},\phi_n,\phi_{n+1}), \quad n \in \mathbb{Z},
$$
where $h : \mathbb{C}^3 \mapsto \mathbb{R}$ is arbitrary function. 
If the non-constant sequence is also real-valued, then $h \equiv 0$. 
\end{Proof}

\begin{Remark}
{\rm The one-to-one correspondence between the second-order and 
first-order difference equations (\ref{second-order-difference}) and 
(\ref{first-order-difference}) exists only for non-constant 
real-valued solutions. Complex-valued solutions of the first-order 
equation (\ref{first-order-difference}) can give solutions of the 
second-order equation with an additional function 
$h(\phi_{n-1},\phi_n,\phi_{n+1})$. This property can be used for a 
full time-space discretization of the NLS equation (\ref{NLS}), when 
the derivative term $\phi'(z)$ in the differential advance-delay 
equation (\ref{advance-delay}) is replaced by the difference term 
$(\phi_{n+1}-\phi_{n-1})/(2h)$ such that the function 
$h(\phi_{n-1},\phi_n,\phi_{n+1}) = c/h$ is constant. We will not 
consider full time-space discretizations of the NLS equation 
(\ref{NLS}) in this paper. }
\end{Remark}

\begin{Lemma}
The reduction of the second-order difference equation
(\ref{second-order-difference}) to the first-order difference
equation (\ref{first-order-difference}) exists provided that the
nonlinear function (\ref{nonlinearity}) satisfies the constraints:
\begin{equation}
\label{constraints-reductions} \alpha_4 = \alpha_1 - \alpha_6, 
\qquad \alpha_5 = \alpha_6, \qquad \alpha_7 = \alpha_1 - 2 \alpha_6, 
\qquad \alpha_{10} = \alpha_8 - \alpha_9,
\end{equation}
such that $(\alpha_1,\alpha_2,\alpha_3,\alpha_6,\alpha_8,\alpha_9) 
\in \mathbb{R}^6$ are free parameters.  \label{lemma-1}
\end{Lemma}

\begin{Proof}
By using symbolic computations with Wolfram's Mathematica we find 
from (\ref{nonlinearity}), (\ref{function-g}) and 
(\ref{correspondence-integral}) that $\alpha_1 = 2 \gamma_1$, 
$\alpha_2 = \gamma_2$, $\alpha_3 = 2 \gamma_3$, and $\alpha_8 = 
\gamma_4$ under the constraints (\ref{constraints-reductions}).
\end{Proof}

\begin{Corollary}
When $\phi_n \in \mathbb{R}$, the first-order difference equation
(\ref{first-order-difference}) is characterized by the symmetric
quartic polynomial function
\begin{equation}
\label{nonlinearity-real}  g = \beta_1 \phi_n^2 \phi_{n+1}^2 +
\beta_2 (\phi_n^2 + \phi_{n+1}^2) \phi_n \phi_{n+1} + \beta_3
(\phi_n^4 + \phi_{n+1}^4),
\end{equation}
where $\beta_1 = \gamma_2 + 2 \gamma_3 = \alpha_2 + \alpha_3$, 
$\beta_2 = 2 \gamma_1 = \alpha_1$, and $\beta_3 = \gamma_4 = 
\alpha_8$ under the continuity constraint
\begin{equation}
\label{continuity-coefficients} \beta_1 + 2 \beta_2 + 2 \beta_3 = 1.
\end{equation}
\label{corollary-0}
\end{Corollary}

\begin{Remark}
{\rm Corollary \ref{corollary-0} recovers the result of
\cite{BOP05}, where real-valued nonlinear functions of the
difference equation (\ref{second-order-difference}) are considered 
in the context of kink solutions of the discrete $\phi^4$ model. }
\end{Remark}

\begin{Lemma}
The discrete NLS equation (\ref{dNLS}) conserves the momentum
invariant (\ref{momentum}) provided that the constraints
(\ref{constraints-reductions}) are met.
 \label{lemma-3}
\end{Lemma}

\begin{Proof}
Computing time-derivative of (\ref{momentum}) and using the
stroboscopical summation in $n$, we convert the result to the
following irreducible remainder:
\begin{eqnarray*}
\dot{M} & = & ( \alpha_1 - \alpha_4 - \alpha_6) \sum_{n \in
\mathbb{Z}}  ( |u_{n+1}|^2 - |u_n|^2) (u_{n+1} \bar{u}_{n} +
\bar{u}_{n+1} u_n)) \\ & + & (\alpha_4 - \alpha_5 - \alpha_7)
\sum_{n \in \mathbb{Z}} ( |u_{n+1}|^2 (u_n \bar{u}_{n-1} + \bar{u}_n
u_{n-1}) - |u_{n-1}|^2 (u_n \bar{u}_{n+1} + \bar{u}_n u_{n+1}) ) \\
& + &  (\alpha_5 - \alpha_6 ) \sum_{n \in \mathbb{Z}} ( u_{n+1} u_n
\bar{u}_{n-1}^2 + \bar{u}_{n+1} \bar{u}_n u_{n-1}^2 - u_{n+1}^2
\bar{u}_n \bar{u}_{n-1} - \bar{u}_{n+1}^2 u_n u_{n-1}) \\
& + & (\alpha_8 - \alpha_9 - \alpha_{10}) \sum_{n \in \mathbb{Z}}
(|u_{n+1}|^2 - |u_{n-1}|^2)( u_{n+1} \bar{u}_{n-1} + \bar{u}_{n+1}
u_{n-1}).
\end{eqnarray*}
The constraints (\ref{constraints-reductions}) give the most general
solution of the system of homogeneous linear equations, which
follows from the momentum conservation $\dot{M} = 0$.
\end{Proof}

\begin{Corollary}
There exists a one-to-one correspondence between the set of
nonlinear functions (\ref{nonlinearity}) that conserves the
momentum (\ref{momentum}) and the set of nonlinear functions that
supports the reduction to the first-order difference equation
(\ref{first-order-difference}).  \label{corollary-2}
\end{Corollary}

\begin{Remark}
{\rm Corollary \ref{corollary-2} proves the conjecture of \cite{K03} 
for the case of a discrete NLS equation (\ref{dNLS}) with the cubic 
polynomial function (\ref{nonlinearity}). A similar result for a 
discrete Klein--Gordon equation was obtained in the most general 
case in \cite[Appendix]{K03} by explicit computations.}
\end{Remark}

\begin{Lemma}
The discrete NLS equation (\ref{dNLS}) has the Hamiltonian structure
(\ref{Hamiltonian}) provided that
\begin{equation}
\label{constraints-Hamiltonians} \alpha_2 = 2 \alpha_3 = 2 \alpha_8,
\qquad \alpha_5 = \alpha_7 = \alpha_9 = \alpha_{10} = 0,
\end{equation}
such that $(\alpha_1,\alpha_2,\alpha_4,\alpha_6) \in \mathbb{R}^4$
are free parameters.  \label{lemma-2}
\end{Lemma}

\begin{Proof}
The most general symmetric real-valued gauge-invariant function 
$F(u_n,u_{n+1})$ of the quartic polynomials in the Hamiltonian 
(\ref{Hamiltonian}) takes the form:
\begin{eqnarray*}
F & = & \delta_1 ( |u_n|^2 + |u_{n+1}|^2) (u_n \bar{u}_{n+1} +
\bar{u}_n u_{n+1}) + \delta_2 |u_n|^2 |u_{n+1}|^2\\
& \phantom{t} & + \delta_3 (u_n^2 \bar{u}_{n+1}^2 + \bar{u}_n^2
u_{n+1}^2) + \delta_4 (|u_n|^4 + |u_{n+1}|^4).
\end{eqnarray*}
Due to the symplectic structure (\ref{Hamiltonian}), the potential
function $F(u_n,u_{n+1})$ produces the nonlinear function
$f(u_{n-1},u_n,u_{n+1})$ in the form,
\begin{eqnarray*}
f & = & \delta_1 \left[  |u_{n+1}|^2 u_{n+1} + |u_{n-1}|^2 u_{n-1} +
2 |u_n|^2 (u_{n+1}+u_{n-1}) + u_n^2 ( \bar{u}_{n+1} + \bar{u}_{n-1})\right] \\
& \phantom{t} & + \delta_2 (|u_{n+1}|^2 + |u_{n-1}|^2 ) u_n + 2
\delta_3 (u_{n+1}^2 + u_{n-1}^2) \bar{u}_n + 4 \delta_4 |u_n|^2 u_n.
\end{eqnarray*}
Therefore, $\alpha_2 = 2 \delta_1$, $\alpha_4 = \delta_2$,
$\alpha_6 = 2 \delta_3$ and $\alpha_1 = 4 \delta_4$ under the
constraints (\ref{constraints-Hamiltonians}).
\end{Proof}

\begin{Corollary}
None of the discrete NLS equations with the Hamiltonian structure 
(\ref{Hamiltonian}) is momentum-preserving, including the dNLS 
lattice (\ref{onsiteNLS}) and the FPU lattice (\ref{FPU}). 
\label{corollary-1}
\end{Corollary}

\begin{Remark}
{\rm Corollary \ref{corollary-1} agrees with the conclusion of
\cite{DKY05} obtained for the discrete $\phi^4$ model.}
\end{Remark}

\begin{Remark}
{\rm The three-parameter discrete NLS equation with the Hamiltonian 
structure (\ref{Hamiltonian}) was derived in \cite{OJE03} for 
modeling of arrays of optical waveguides. The model of \cite{OJE03} 
corresponds to the reduction of the symmetric potential function 
$F(u_n,u_{n+1})$ with $\delta_2 = 4 \delta_3$. Although it is 
claimed in \cite{OJE03} from results of numerical simulations that 
traveling solutions may have enhanced mobility in this model, no 
momentum-preservation is supported and the second-order difference 
equation (\ref{second-order-difference}) provides no reduction to 
the first-order difference equation (\ref{first-order-difference}).}
\end{Remark}

\section{Existence of translationally invariant stationary solutions}

We shall consider existence of continuous solutions of the 
advance-delay equation (\ref{advance-delay}) with $c = 0$ from 
existence of the translationally invariant solutions of the 
second-order difference equation (\ref{second-order-difference}). We 
show that there exist translationally invariant stationary solutions 
if the second-order difference equation 
(\ref{second-order-difference}) is reduced to the first-order 
difference equation (\ref{first-order-difference}). Thus, the 
existence of the reduction to the first-order difference equation 
(\ref{first-order-difference}) gives the {\em sufficient} condition 
for existence of translationally invariant stationary solutions. For 
simplicity, the analysis is developed for {\em real-valued} 
solutions of the first-order difference equation 
(\ref{first-order-difference}). By Lemma 
\ref{lemma-correspondence-equations}, such non-constant solutions 
correspond to real-valued solutions of the second-order difference 
equation (\ref{second-order-difference}). The initial-value problem 
for the first-order difference equation 
(\ref{first-order-difference}) with $E = 0$ in the space of 
real-valued localized solutions takes the implicit form,
\begin{equation}
\label{first-order-map} \left\{ \begin{array}{l} (\phi_{n+1} - 
\phi_n )^2 = h^2 \omega \phi_n \phi_{n+1} -
h^2 g(\phi_n,\phi_{n+1}), \qquad n \in \mathbb{Z},   \\
\phi_0 = \varphi,  \end{array} \right.
\end{equation}
where $\varphi \in \mathbb{R}$ is the initial data, iterations in 
both positive and negative directions of $n \in \mathbb{Z}$ are 
considered, and $g : \mathbb{R}^2 \mapsto \mathbb{R}$ is given by 
(\ref{nonlinearity-real}). Let $x = \phi_n$ and $y = \phi_{n+1}$ and 
rewrite $g(\phi_n,\phi_{n+1})$ as
$$
g(x,y) = \beta_1 x^2 y^2 + \beta_2 x y (x^2 + y^2)  + \beta_3 (x^4 + 
y^4).
$$
Due to the continuity constraint (\ref{continuity-coefficients}), we 
have $g(x,x) = x^4$ and $\partial_x g(x,x) = \partial_y g(x,x) = 2 
x^3$. We first establish the sufficient condition that the 
second-order difference equation (\ref{second-order-difference}) has 
only real-valued solutions and then develop analysis of the 
initial-value problem (\ref{first-order-map}) in $n \in \mathbb{Z}$. 
Our main result is Proposition \ref{theorem-existence} which 
guarantees  existence of a particular single-humped localized 
sequence $\{ \phi_n \}_{n \in \mathbb{Z}}$ to the initial-value 
problem (\ref{first-order-map}) with $\omega > 0$ and sufficiently 
small $h$. The sequence corresponds to the translationally invariant 
solution of the second-order equation 
(\ref{second-order-difference}) which converges as $h \to 0$ to the 
solution $\phi_s(z) = \sqrt{\omega} \; {\rm sech} (\sqrt{\omega} z)$ 
at the points $z = hn$ with $n \in \mathbb{Z}$. 

\begin{Lemma}
The second-order difference equation (\ref{second-order-difference}) 
with sufficiently small $h$ admits a conserved quantity 
\begin{equation}
\label{flux} J = \bar{\phi}_n \phi_{n+1} - \phi_n \bar{\phi}_{n+1}
\end{equation}
for bounded sequences $\{ \phi_n \}_{n \in \mathbb{Z}}$ if
\begin{equation}
\label{constraints-realvalued} \alpha_7 = 2 \alpha_6, \qquad
\alpha_{10} = \alpha_8 + \alpha_9.
\end{equation}
\label{lemma-10}
\end{Lemma}

\begin{Proof}
The nonlinear function (\ref{nonlinearity}) satisfies the relation 
{\small
\begin{eqnarray*}
\bar{\phi}_n f(\phi_{n-1},\phi_n,\phi_{n+1}) - \phi_n
\overline{f(\phi_{n-1},\phi_n,\phi_{n+1})} =  \left( \bar{\phi}_n 
\phi_{n+1} - \phi_n \bar{\phi}_{n+1} + \bar{\phi}_n \phi_{n-1} - 
\phi_n \bar{\phi}_{n-1}\right) Q_n + R_n
\end{eqnarray*}
}where
\begin{eqnarray*}
Q_n = (\alpha_2 - \alpha_3) |\phi_n|^2 + \alpha_8 ( |\phi_{n-1}|^2
+ |\phi_{n+1}|^2) + \alpha_9 (\bar{\phi}_{n-1}
\phi_{n+1} + \phi_{n-1} \bar{\phi}_{n+1}) \\
+ \alpha_6 ( \bar{\phi}_n \phi_{n+1} + \phi_n \bar{\phi}_{n+1} +
\bar{\phi}_n \phi_{n-1} + \phi_n \bar{\phi}_{n-1} )
\end{eqnarray*}
and
\begin{eqnarray*}
R_n = (\alpha_7 - 2 \alpha_6) (\bar{\phi}_n^2 \phi_{n-1}
\phi_{n+1} -
\phi_n^2 \bar{\phi}_{n-1} \bar{\phi}_{n+1} ) \\
+ (\alpha_{10} - \alpha_8 - \alpha_9) \left[ (\bar{\phi}_n 
\phi_{n-1} - \phi_n \bar{\phi}_{n-1}) |\phi_{n+1}|^2 + (\bar{\phi}_n 
\phi_{n+1} - \phi_n \bar{\phi}_{n+1}) |\phi_{n-1}|^2 \right].
\end{eqnarray*}
If $\alpha_7 = 2 \alpha_6$ and $\alpha_{10} = \alpha_8 + \alpha_9$, 
then $R_n \equiv 0$. If the sequence $\{ \phi_n \}_{n\in 
\mathbb{Z}}$ is bounded, then $1 + h^2 Q_n > 0$ and $J$ is constant 
for all $n \in \mathbb{Z}$ provided that $h$ is sufficiently small. 
\end{Proof}

\begin{Corollary}
Under the constraints (\ref{constraints-realvalued}), real-valued 
solutions of the second-order difference equation 
(\ref{second-order-difference}) with $J = 0$ are the only localized 
bounded solutions $\{ \phi_n \}_{n \in \mathbb{Z}}$ with $\phi_0 \in 
\mathbb{R}$. \label{corollary-8}
\end{Corollary}

\begin{Remark}
{\rm While it is not clear if localized complex-valued solutions of 
the second-order difference equation (\ref{second-order-difference}) 
may exist when $\alpha_7 \neq 2 \alpha_6$ and $\alpha_{10} \neq 
\alpha_8 + \alpha_9$, we observe that the reduction to the 
first-order difference equation (\ref{first-order-difference}) has 
six free parameters (see Lemma \ref{lemma-1}) but the nonlinear 
function in the form (\ref{function-g}) has only four parameters. 
Therefore, two parameters can be used to satisfy the constraints 
(\ref{constraints-realvalued}). Combining the constraints 
(\ref{constraints-reductions}) and (\ref{constraints-realvalued}), 
we obtain the most general parameterization of the nonlinear 
function (\ref{nonlinearity}) that admits the conserved quantities 
(\ref{first-order-difference}) and (\ref{flux}):
\begin{equation}
\label{constraints-final} \alpha_1 = 4 \alpha_6, \quad \alpha_4 =
3 \alpha_6, \quad \alpha_5 = \alpha_6, \quad \alpha_7 = 2
\alpha_6, \quad \alpha_9 = 0, \quad \alpha_{10} = \alpha_8,
\end{equation}
where $(\alpha_2,\alpha_3,\alpha_6,\alpha_8) \in \mathbb{R}^4$ are 
free parameters. By Lemma \ref{lemma-correspondence-equations} and 
Corollary \ref{corollary-8}, all localized non-constant solutions of 
the second-order difference equation (\ref{second-order-difference}) 
under the constraints (\ref{constraints-final}) are real-valued and 
the sequence $\{ \phi_n \}_{n \in \mathbb{Z}}$ is equivalently found 
from the initial-value problem (\ref{first-order-map}).} 
\end{Remark}

\begin{figure}[htbp]
\begin{center}
\includegraphics[height=7cm]{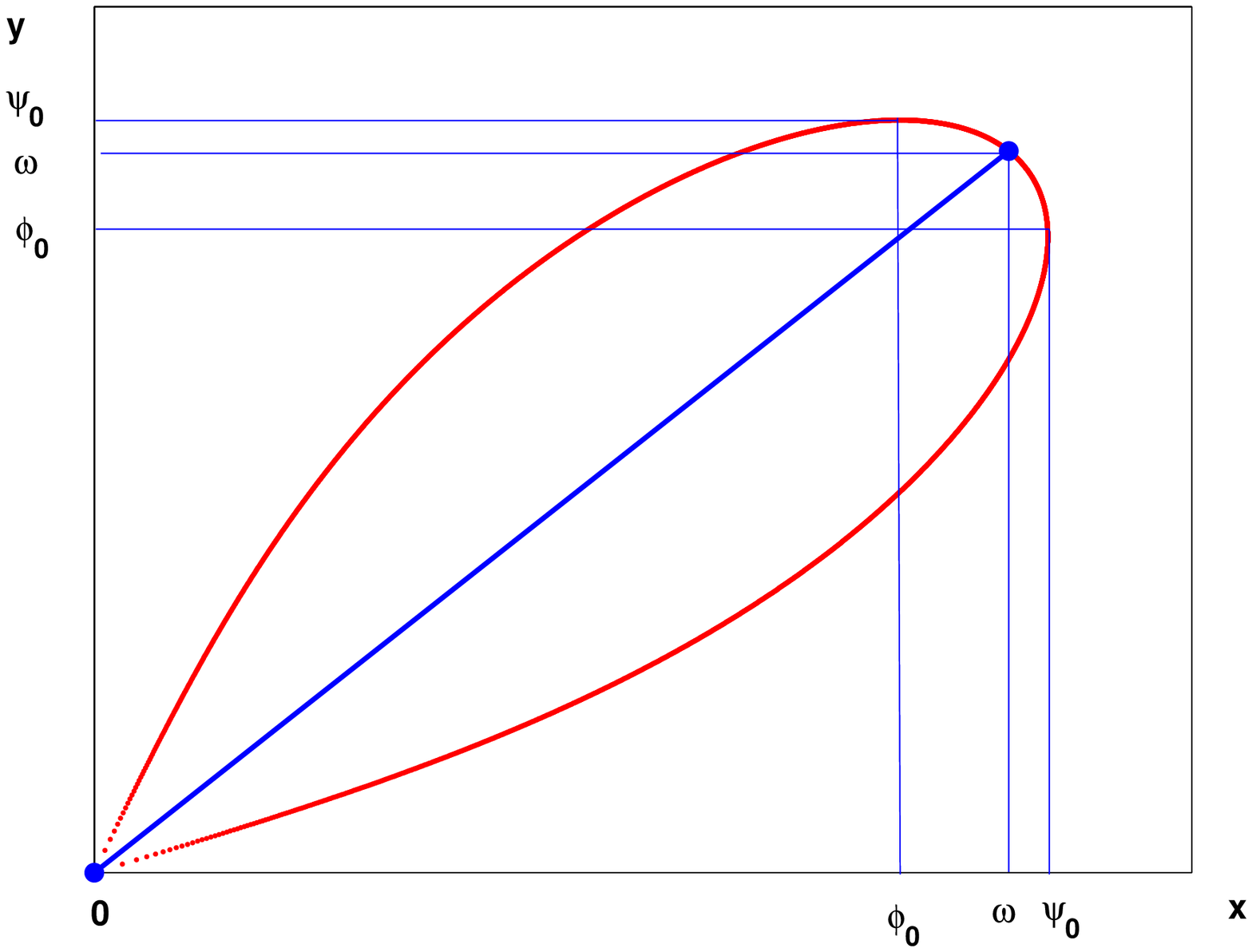}  
\end{center}
\caption{Solutions of the quartic equation (\ref{two-roots}) with 
$\beta_1 = \beta_2 = 0$ and $\beta_3 = 0.5$. }
\end{figure}

\begin{Lemma}
Let $Q(x,y) = \omega x y - g(x,y)$ and $\omega > 0$ is parameter.
There exists $h_0 > 0$ and $L > \sqrt{\omega}$ such that the
algebraic (quartic) equation
\begin{equation}
\label{two-roots} (y-x)^2 = h^2 Q(x,y)
\end{equation}
defines a convex, simply-connected, closed curve inside the domain 
$0 \leq x \leq L$ and $0 \leq y \leq L$ for $0 < h < h_0$. The curve 
is symmetric about the diagonal $y = x$ and has two intersections 
with the diagonal at $y = x = 0$ and $y = x = \sqrt{\omega}$. 
\label{lemma-roots}
\end{Lemma}

\begin{Proof}
The statement of lemma is illustrated on Figure 1. Symmetry of the 
curve about the diagonal $y = x$ follows from the fact that $g(x,y) 
= g(y,x)$. Intersections with the diagonal $y = x$ follows from the 
continuity constraint that gives $Q(x,x) = x^2( \omega - x^2)$. To 
prove convexity, we consider the behavior of the curve for 
sufficiently small $h$ in two domains: in the interval $0 \leq x 
\leq x_0$ and in a small neighborhood of the point $y = x = 
\sqrt{\omega}$ with the radius $r_0 h^2$, where $x_0 = \sqrt{\omega} 
- r_0 h^2$ and $r_0 > \frac{\sqrt{\omega^3}}{8}$ is $h$-independent. 
Let $y = x + z$ and consider two branches of the algebraic equation 
(\ref{two-roots}) in the implicit form
\begin{equation}
\label{upper-root} F_{\pm}(x,z,h) = z \mp h \sqrt{Q(x,x+z)} = 0.
\end{equation}
such that for $x \geq 0$
\begin{eqnarray*}
F_{\pm}(x,0,h) = \mp h x \sqrt{\omega - x^2}, \qquad
\partial_z F_{\pm}(x,0,h) = 1 \mp \frac{h (\omega - 2x^2)}{2 \sqrt{\omega - x^2}}.
\end{eqnarray*}
It is clear that $F_{\pm}(x,0,0) = 0$ and $\partial_z F_{\pm}(x,0,0) 
= 1$, while $F_{\pm}(x,0,h)$ is continuously differentiable in $x$ 
and $h$ and $\partial_z F_{\pm}(x,0,h)$ is uniformly bounded in $x$ 
and $h$ on any compact subset of $x \in [0, x_0]$, where $x_0 = 
\sqrt{\omega} - r_0 h^2$ and $r_0 > \frac{\sqrt{\omega^3}}{8}$ is 
$h$-independent. By the Implicit Function Theorem, there exists $h_0 
> 0$, such that the implicit equations (\ref{upper-root}) define unique roots 
$z = \pm h S_{\pm}(x,h)$ in the domain $0 \leq x \leq x_0$ and $0 < 
h < h_0$, where $h S_{\pm}(x,h)$ are positive, continuously 
differentiable functions in $x$ and $h$. Therefore, the algebraic 
equation (\ref{two-roots}) defines two branches of the curve located 
above and below the diagonal $y = x$. When $h_0$ is sufficiently 
small, the two branches are strictly increasing in the interval $0 
\leq x \leq x_0$. In the limit $h_0 \to 0$, the two branches 
converge to the diagonal $y = x$ on $0 \leq x \leq \sqrt{\omega}$. 
Derivatives of the algebraic equation (\ref{two-roots}) in $x$ are 
defined for any branch of the curve by 
\begin{eqnarray}
\label{first-derivative} y' \left[ 2 (y-x) - h^2(\omega x - 
\partial_y g(x,y)) \right] & = & 
2(y - x) + h^2(\omega y - \partial_x g(x,y)), \\
\label{second-derivative} y'' \left[ 2 (y-x) - h^2(\omega x - 
\partial_y g(x,y)) \right] & = & -2(y' - 1)^2 + h^2(2 \omega y' - 
\tilde{g}),
\end{eqnarray}
where $\tilde{g} = \partial^2_{xx} g(x,y) - 2 y' \partial^2_{xy} 
g(x,y) - (y')^2\partial^2_{yy} g(x,y)$. It follows from the first 
derivative (\ref{first-derivative}) that $y' = -1$ at the point $y = 
x = \sqrt{\omega}$ for any $h > 0$, such that there exists a small 
neighborhood of the point $y = x = \sqrt{\omega}$ where the upper 
branch of the curve is strictly decreasing. Let $B_h^+$ be the upper 
semi-disk centered at $y = x = \sqrt{\omega}$ with a radius $r_0 
h^2$, where $r_0 > \frac{\sqrt{\omega^3}}{8}$ is $h$-independent. 
Then, there exists a $h$-independent constant $C > 0$ such that
$$
2 (y-x) - h^2(\omega x - \partial_y g(x,y)) \geq C h^2, \qquad (x,y) 
\in B_h^+.
$$
It follows from the second derivative (\ref{second-derivative}) for 
sufficiently small $h$ that $y'' \leq -C_1/h^2$ in $(x,y) \in B_h^+$ 
with $C_1 > 0$. Therefore, the curve is convex in $B_h^+$. By using 
the rescaled variables in $B_h^+$:
$$
x = \sqrt{\omega} - X h^2, \qquad y = \sqrt{\omega} + Y h^2,
$$
we find that the curvature of the curve in new variables is bounded 
by $Y''(X) \leq - C_1$ and therefore, the first derivative $Y'(X)$ 
and so $y'(x)$ may only change by a finite number in $(x,y) \in 
B_h$. Therefore, there exists $0 < C_2 < \infty$ such that $y'(x) = 
C_2$ at a point $x$, where $x \leq x_0 = \sqrt{\omega} - r_0 h^2$. 
By the first part of the proof, the upper branch of the curve is 
monotonically increasing for $0 \leq x \leq x_0$. By the second part 
of the proof, it has a single maximum for $x_0 \leq x \leq 
\sqrt{\omega}$. Thus, the curve defined by the quartic equation 
(\ref{two-roots}) is convex for $y \geq x \geq 0 $ (and for $0 \leq 
y \leq x$ by symmetry). 
\end{Proof}

\begin{Corollary}
Let $\psi_0$ be a maximal value of $y$ and $\varphi_0$ be the 
corresponding value of $x$ on the upper branch of the curve defined 
by the quartic equation (\ref{two-roots}). Then, $\varphi_0 < 
\sqrt{\omega} < \psi_0$ and 
\begin{equation}
\label{asumptotic-maximum} \varphi_0 = \sqrt{\omega} - \frac{3}{8} 
\sqrt{\omega^3} h^2 + {\rm O}(h^4), \qquad \psi_0 = \sqrt{\omega} + 
\frac{1}{8} \sqrt{\omega^3} h^2 + {\rm O}(h^4). 
\end{equation}
\end{Corollary}

\begin{Lemma}
The initial-value problem (\ref{first-order-map}) with $\omega > 0$ 
and $0 < h < h_0$ admits a unique monotonically decreasing sequence 
$\{ \phi_n \}_{n = 0}^{\infty}$ for any $0 < \varphi < 
\sqrt{\omega}$ that converges to zero from above as $n \to \infty$. 
The sequence $\{ \phi_n \}_{n=0}^{\infty}$ is continuous with 
respect to $h$ and $\varphi$. \label{lemma-increasing}
\end{Lemma}

\begin{Proof}
By Lemma \ref{lemma-roots} (see Figure 1), there exists a unique 
lower branch of the curve in (\ref{two-roots}) below the diagonal $y 
= x$ for $0 < x < \sqrt{\omega}$ and the monotonically decreasing 
sequence $\{ \phi_n \}_{n = 0}^{\infty}$ with $0 < \varphi < 
\sqrt{\omega}$ satisfies the initial-value problem:
\begin{equation}
\label{IVP-linear} \left\{ \begin{array}{l} \phi_{n+1} = \phi_n - h
S_-(\phi_n,h), \qquad n \in \mathbb{N}  \\ \phi_0 = \varphi
\end{array} \right.
\end{equation}
where $h S_-(\phi,h) > 0$ is continuously differentiable with 
respect to $\phi$ and $h$. We shall prove that the monotonically 
decreasing sequence $\{ \phi_n \}_{n=0}^{\infty}$ converges to zero 
from above. Since $Q(x,y)$ is a quartic polynomial, there exists a 
constant $C > 0$ that depends on $\omega$ and is independent of $h$, 
such that
\begin{equation}
\label{decay-estimate} (\phi_{n+1} - \phi_n)^2 \leq C h^2 \phi_n^2
\end{equation}
for all $\phi_{n+1} < \phi_n$. If $h$ is sufficiently small, such 
that $C h^2 < 1$, then $0 < \phi_{n+1} < \phi_n$, and the sequence 
$\{\phi_n\}_{n=0}^{\infty}$ is bounded from below by $\phi = 0$. By 
the Weierstrass Theorem, the monotonically decreasing and bounded 
from below sequence $\{ \phi_n \}_{n=0}^{\infty}$ converges as $n 
\to \infty$ to the fixed point $\phi = 0$. Continuity of the 
sequence $\{ \phi_n \}_{n=0}^{\infty}$ in $h$ and $\varphi$ follows 
from smoothness of $h S_-(\phi,h)$ in $h$ and $\varphi$.
\end{Proof}

\begin{Lemma}
There exists $N \geq 1$, such that the initial-value problem 
(\ref{first-order-map}) with $\omega > 0$ and $0 < h < h_0$ admits a 
unique monotonically increasing sequence $\{ \phi_n \}_{n = 
-\infty}^N$ for any $0 < \varphi < \sqrt{\omega}$ that converges to 
zero from above as $n \to -\infty$. The sequence $\{ \phi_n 
\}_{n=-\infty}^N$ is continuous with respect to $h$ and $\varphi$. 
\label{lemma-decreasing}
\end{Lemma}

\begin{Proof}
By Lemma \ref{lemma-roots} (see Figure 1), there exists a unique 
upper branch of the curve in (\ref{two-roots}) above the diagonal $y 
= x$ for $0 < x < \sqrt{\omega}$ and the monotonically increasing 
sequence $\{ \phi_n \}_{n = -\infty}^0$ with $0 < \varphi < 
\sqrt{\omega}$ satisfies the initial-value problem:
\begin{equation}
\label{IVP-linear-2} \left\{ \begin{array}{l} \phi_{n+1} = \phi_n + 
h S_+(\phi_n,h), \qquad (-n) \in \mathbb{N}  \\ \phi_0 = \varphi
\end{array} \right.
\end{equation}
where $h S_+(\phi,h) > 0$ is continuously differentiable with 
respect to $\phi$ and $h$. Existence of $N \geq 1$ follows from the 
same equation (\ref{IVP-linear-2}) for $0 \leq n \leq N-1$. The 
proof that the monotonically increasing sequence $\{ \phi_n 
\}_{n=-\infty}^{N}$ converges to zero from above as $n \to -\infty$ 
is similar to the proof of Lemma \ref{lemma-increasing}. Continuity 
of the sequence $\{ \phi_n \}_{n=-\infty}^{N}$ in $h$ and $\varphi$ 
follows from smoothness of $h S_+(\phi,h)$ in $h$ and $\varphi$.
\end{Proof}

\begin{Lemma}
The initial-value problem (\ref{first-order-map}) with $\omega > 0$ 
and $0 < h < h_0$ admits a unique 2-periodic orbit $\{ \phi_n \}_{n 
\in \mathbb{Z}}$ with $\phi_{n+1} \neq \phi_n$ and $\phi_{n+2} = 
\phi_n$ for any $\varphi_0 < \varphi < \psi_0$.
\end{Lemma}

\begin{Proof}
By Lemma \ref{lemma-roots} (see Figure 1), the curve in 
(\ref{two-roots}) is symmetric about $y = x$ and has two branches in 
$x$ below $y = x$ for $\sqrt{\omega} < x < \psi_0$. Therefore, any 
initial data on the branch between $(\varphi_0,\psi_0)$ and 
$(\psi_0,\varphi_0)$ leads to a unique 2-periodic orbit.
\end{Proof}

\begin{Corollary}
The initial-value problem (\ref{first-order-map}) with $\omega > 0$ 
and $0 < h < h_0$ admits the following particular sequences:
\begin{itemize}
\item For any given $0 < \varphi \leq \psi_0$, there exists a
symmetric single-humped sequence $\{ \phi_n \}_{n\in \mathbb{Z}}$ 
with maximum at $n = 0$, such that $\phi_{n} = \phi_{-n}$ (see 
Figure 2(a)). The single-humped sequence is unique for $\varphi = 
\psi_0$ (see Figure 2(b)). Let $S_{\rm on}$ denote the countable 
infinite set of values of $\{ \phi_n \}_{n \in \mathbb{Z}}$ for the 
single-humped solution with $\varphi = \psi_0$.

\item For any given $\sqrt{\omega} < \varphi < \psi_0$, there exists a 
symmetric double-humped sequence $\{ \phi_n \}_{n\in \mathbb{Z}}$ 
with local minimum at $n = 1$ and maxima at $n = 0$ and $n = 2$, 
such that $\phi_{n} = \phi_{-n+2}$ and $\varphi_0 < \phi_1 <
\sqrt{\omega}$ (see Figure 3(a)). The double-humped sequence becomes 
a unique $2$-site top single-humped sequence for $\varphi = 
\sqrt{\omega}$ (see Figure 3(b)). Let $S_{\rm off}$ denote the 
countable infinite set of values of $\{ \phi_n \}_{n \in 
\mathbb{Z}}$ for the $2$-site top single-humped solution with 
$\varphi = \sqrt{\omega}$. 

\item For any $\varphi \in (0,\psi_0) \backslash \{ S_{\rm on},S_{\rm off}\}$, 
there exists a unique non-symmetric single-humped sequence $\{ 
\phi_n \}_{n\in \mathbb{Z}}$ with $\phi_k \neq \phi_m$ for all $k 
\neq m$.
\end{itemize}
\label{corollary-4}
\end{Corollary}

\begin{figure}[htbp]
\begin{center}
\includegraphics[height=5.5cm]{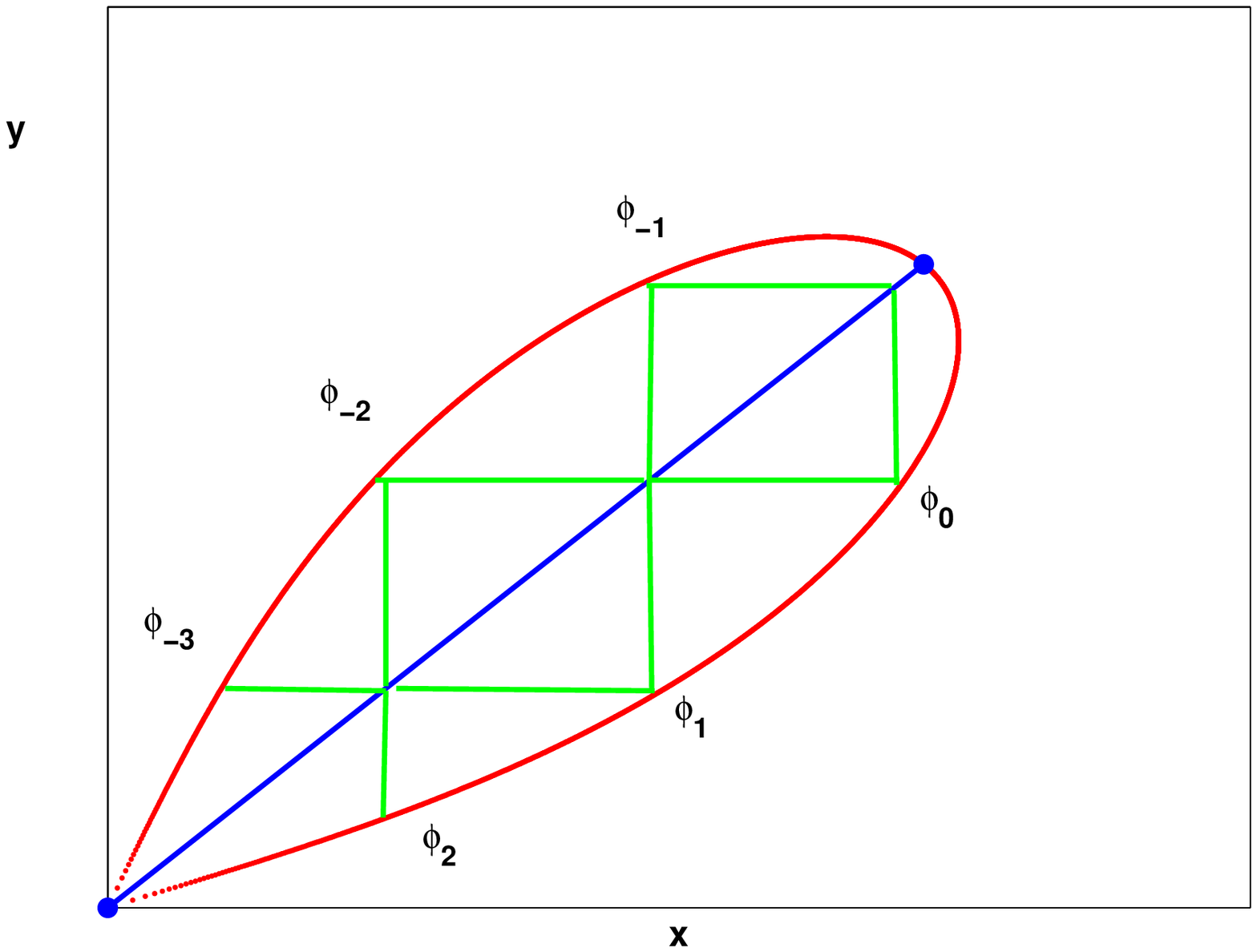} 
\includegraphics[height=5.5cm]{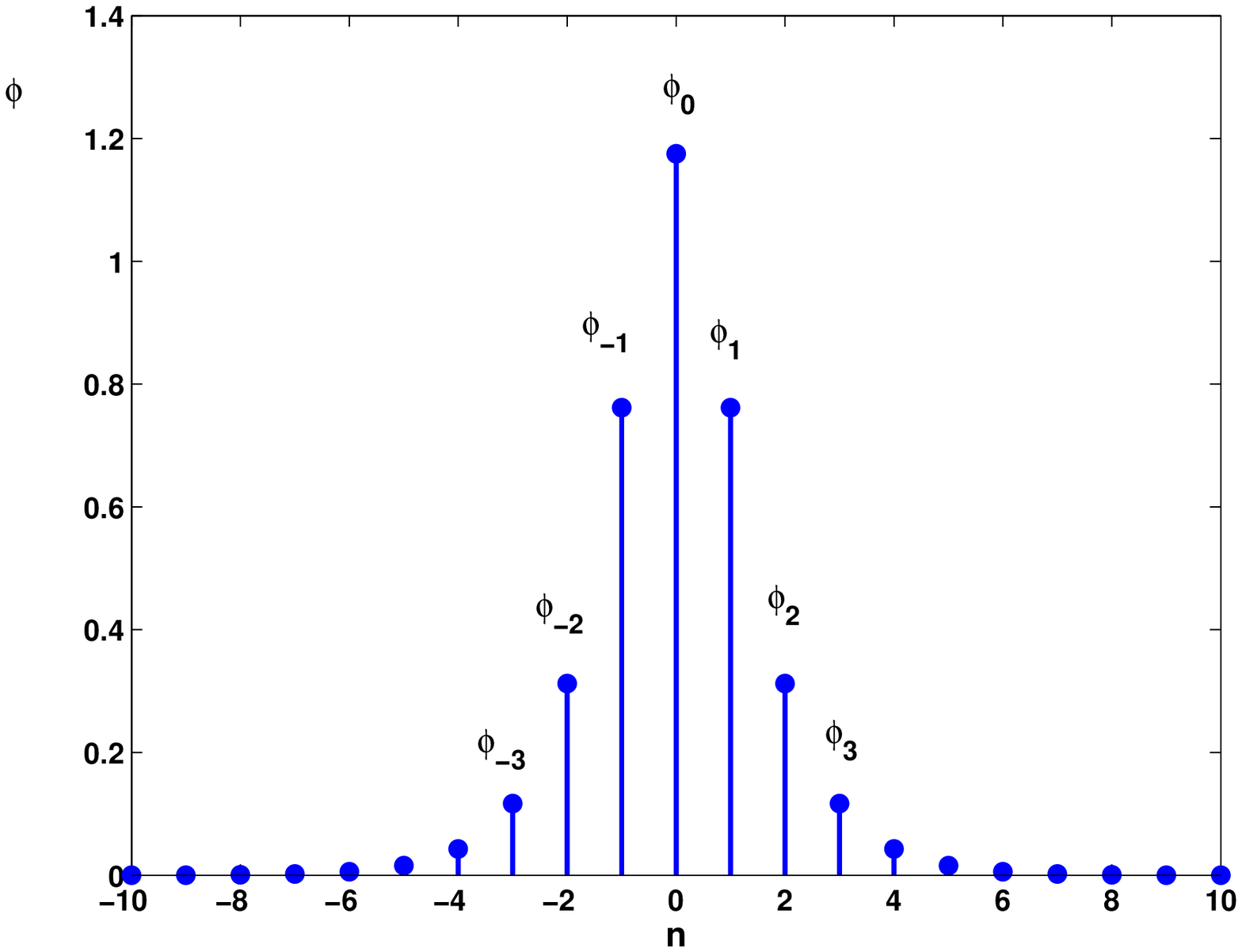}  
\end{center}
\caption{Left: a construction of the symmetric single-humped 
sequence from the solution of the quartic equation 
(\ref{two-roots}). Right: an example of the symmetric single-humped 
solution of the difference equation (\ref{first-order-map}) for 
$\varphi = \psi_0$.}
\end{figure}

\begin{figure}[htbp]
\begin{center}
\includegraphics[height=5.5cm]{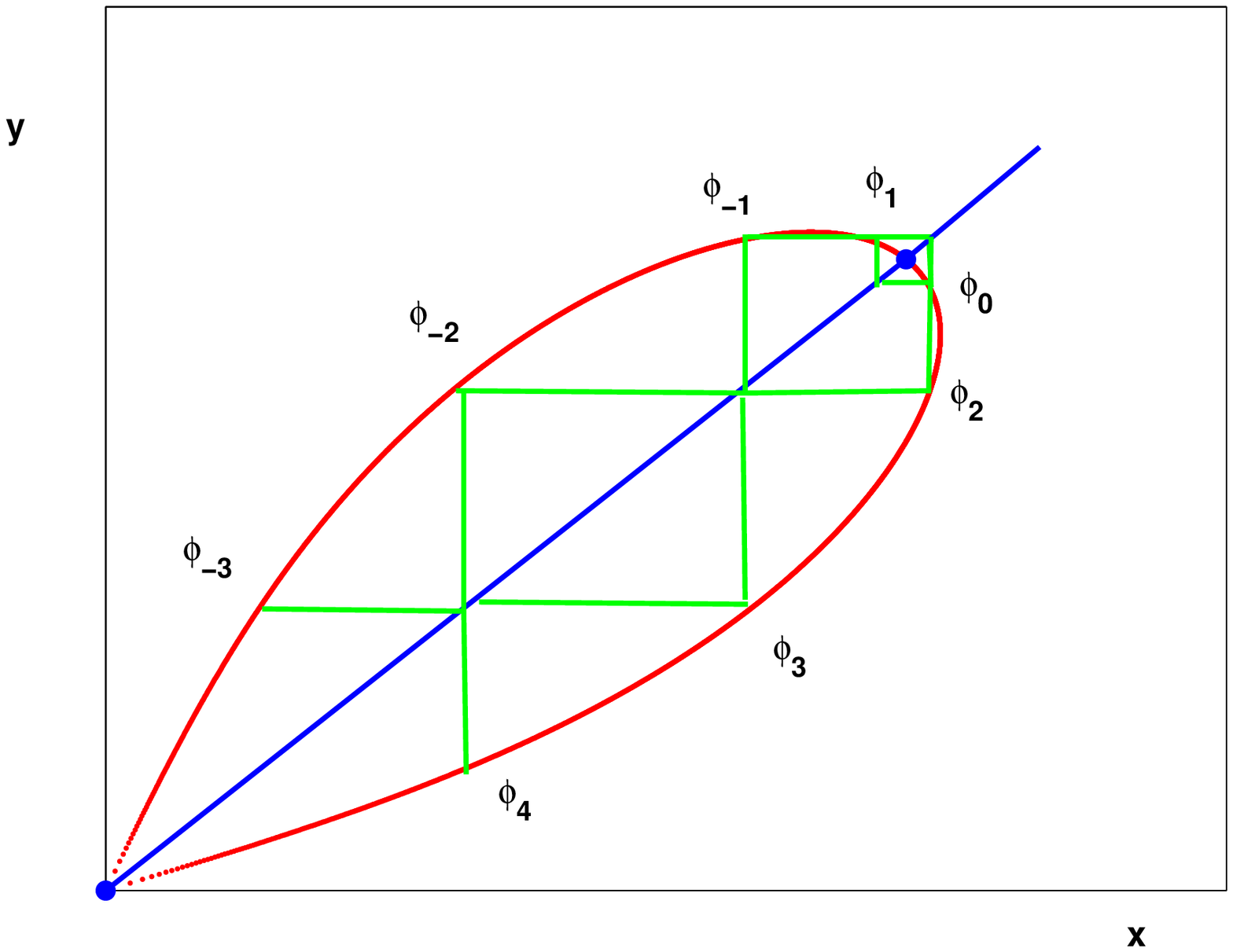} 
\includegraphics[height=5.5cm]{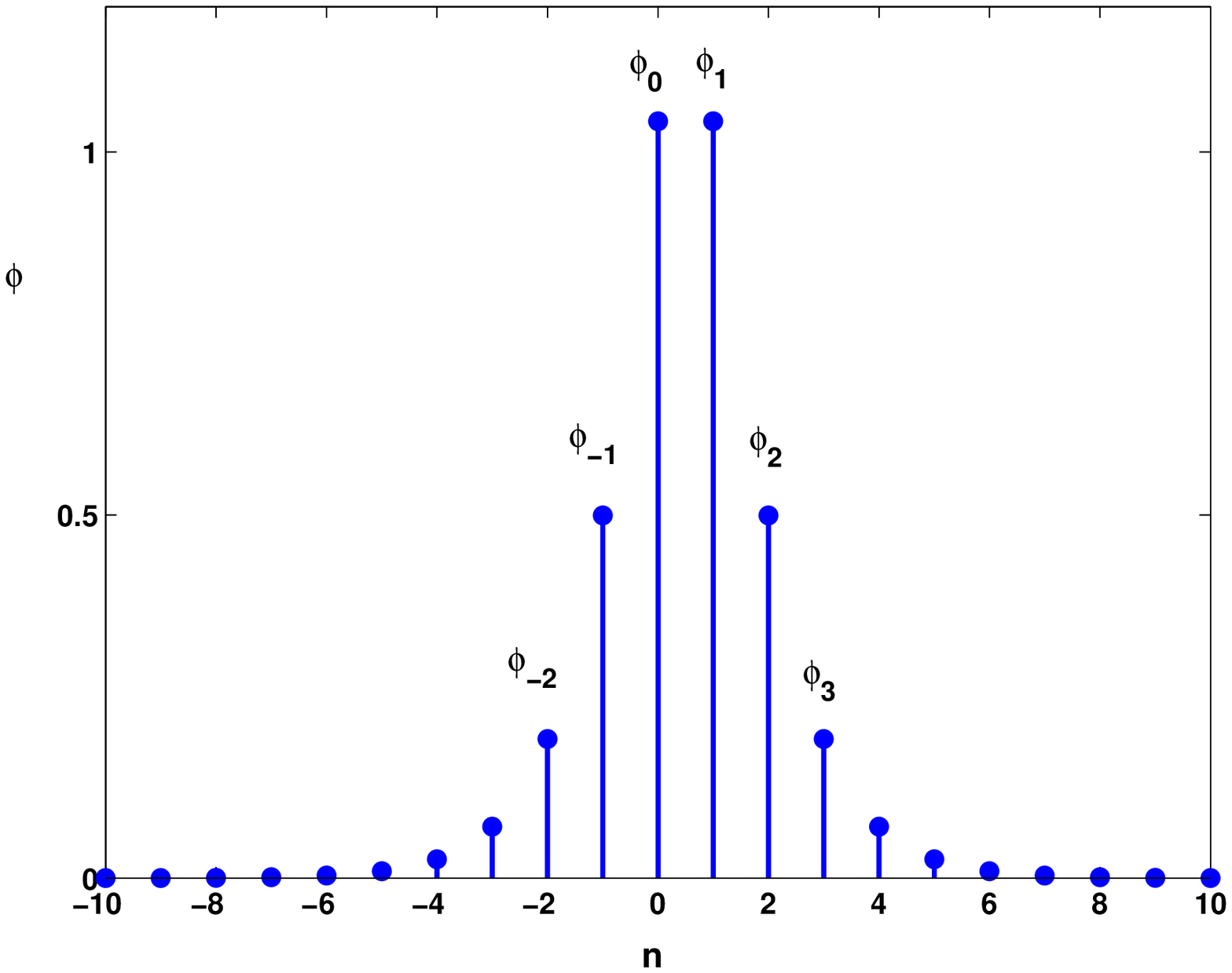}  
\end{center}
\caption{Left: a construction of the symmetric double-humped 
sequence from the solution of the quartic equation 
(\ref{two-roots}). Right: an example of the 2-site top single-humped 
solution of the difference equation (\ref{first-order-map}) for 
$\varphi = \sqrt{\omega}$.}
\end{figure}

\begin{Proposition}
The second-order difference equation (\ref{second-order-difference}) 
admits a translationally invariant single-humped  sequence $\{ 
\phi_n \}_{n\in \mathbb{Z}}$ for any $0 < \phi_0 \leq \psi_0$ with 
$\phi_n = \phi(nh-s)$, where $n \in \mathbb{Z}$, $s \in \mathbb{R}$, 
and $\phi(z)$ is a continuous function. In the limit $h \to 0$, the 
function $\phi(z)$ converges pointwise to the function $\phi_s(z) = 
\sqrt{\omega} \; {\rm sech}(\sqrt{\omega} z)$, i.e. there exists $C 
> 0$ and $s \in \mathbb{R}$ such that 
\begin{equation}
\label{error-bound-discrete-space} \max_{n \in \mathbb{Z}} |\phi_n - 
\phi_s(nh - s)| \leq C h^2.
\end{equation}
\label{theorem-existence}
\end{Proposition}

\begin{Proof}
By the geometric construction of Corollary \ref{corollary-4}, there 
exists a single-humped sequence $\{ \phi_n \}_{n\in \mathbb{Z}}$ for 
any $0 < \phi_0 \leq \psi_0$, which consists of the unique symmetric 
sequence for $\phi_0 \in S_{\rm on}$, the unique $2$-site top 
sequence for $\phi_0 \in S_{\rm off}$ and the unique non-symmetric 
sequence for $\phi_0 \in (0,\psi_0) \backslash \{ S_{\rm on},S_{\rm 
off} \}$. By Lemmas \ref{lemma-increasing} and 
\ref{lemma-decreasing}, the single-humped sequence is continuous in 
$h$ and $\phi_0$ such that it is translationally invariant. By Lemma 
\ref{lemma-correspondence-equations}, the real-valued non-constant 
sequence satisfies identically the second-order equation 
(\ref{second-order-difference}). It remains to show the pointwise 
convergence of the sequence $\{ \phi_n \}_{n\in \mathbb{Z}}$ to the 
sequence $\{ \phi_s(nh - s) \}_{n \in \mathbb{Z}}$ as $h \to 0$, 
where $\phi_s(z) = \sqrt{\omega} \; {\rm sech}(\sqrt{\omega} z)$. By 
the translational invariance of the sequence, we can place the 
maximum at $n = 0$, such that $\phi_0 = \psi_0$, which is equivalent 
to the choice $s = 0$. Then, the bound 
(\ref{error-bound-discrete-space}) with $s = 0$ coincides with the 
bound for the convergence of the symmetric single-humped solution 
with $\{ \phi_n \}_{n \in \mathbb{Z}} = S_{\rm on}$ of the 
second-order difference equation (\ref{second-order-difference}) to 
the solution $\phi_s(hn) = \sqrt{\omega} \; {\rm sech}(\sqrt{\omega} 
h n)$ of the second-order ODE (\ref{second-order-ODE}). The standard 
proof of the error bound (\ref{error-bound-discrete-space}) is 
performed with a formal power series in $h^2$ which involves 
hyperbolic functions for exponentially decaying solutions (see 
Appendix B in \cite{MacKay}). The formal series is different from 
the rigorous convergent asymptotic solution of the difference map 
(\ref{second-order-difference}) along the stable and unstable 
manifolds by the exponentially small in $h$ terms (see \cite{Tovbis} 
for application of this technique to the second-order difference 
equation with a quadratic nonlinearity). We note that a similar 
result (but with a different technique based on analysis of Fourier 
transforms) was proved in \cite{FP99} for the discrete 
Fermi--Pasta--Ulam problem.   
\end{Proof}

\begin{Example}
{\rm Let us consider the explicit example of the AL lattice 
(\ref{AL}), when $\beta_1 = 1$ and $\beta_2 = \beta_3 = 0$. In this 
case, the second-order difference equation 
(\ref{second-order-difference}) admits the exact solution for 
translationally invariant single-humped solutions:
\begin{equation}
\label{soliton-exact} \phi_n = \psi_0 \; {\rm sech}(\kappa h n - s),
\end{equation}
where $(s,\kappa) \in \mathbb{R}^2$ are arbitrary and 
$(\psi_0,\omega)$ are defined by 
\begin{equation}
\label{parameter-exact} \psi_0 = \frac{\sinh(\kappa h)}{h},\qquad 
\omega = \frac{4}{h^2} \sinh^2 \left(\frac{\kappa h}{2}\right).
\end{equation}
The set $S_{\rm on}$ for the symmetric single-humped solution is 
defined by values of $\{\phi_n\}_{n \in \mathbb{Z}}$ for $s = 0$ 
(see Figure 2(b)) and the set $S_{\rm off}$ for the 2-site top 
single-humped solution is defined for $s = \frac{\kappa h}{2}$. We 
find that 
\begin{equation}
\label{value-psi-0} \psi_0 = \sqrt{\omega} \cosh\left( \frac{\kappa 
h}{2}\right) = \sqrt{\omega + \frac{\omega^2 h^2}{4}},
\end{equation}
in agreement with the asymptotic formula (\ref{asumptotic-maximum}). 
The algebraic equation (\ref{two-roots}) can be reduced to the 
explicit roots for the one-step maps $h S_{\pm}(\phi,h)$ in the 
form:
$$
S_{\pm}(\phi,h) = \frac{\phi ( \sqrt{4(\omega - \phi^2) + h^2 
\omega^2} \pm h(\omega - 2 \phi^2))}{2(1 + h^2 \phi^2)},
$$
which satisfy all properties derived in Lemma \ref{lemma-roots}. In 
particular $S_+(\phi,h) > 0$ for $0 < \phi < \sqrt{\omega}$ and 
$S_-(\phi,h) > 0$ has one root for $0 < \phi < \sqrt{\omega}$ and 
two roots for $\sqrt{\omega} < \phi < \psi_0$, where $\psi_0$ is 
given by (\ref{value-psi-0}). In the limit $h \to 0$, the expression 
(\ref{soliton-exact}) with (\ref{parameter-exact}) converges to the 
solution (\ref{soliton}) with $\omega = \kappa^2$ and $c = 0$. }
\end{Example}

\begin{Remark} 
{\rm For real-valued non-constant solutions, the second-order 
difference equation (\ref{second-order-difference}) can be reduced 
to another first-order difference equation:
\begin{equation}
\label{first-order-linear} \frac{1}{h} (\phi_{n+1}-\phi_n) =
\tilde{g}(\phi_n,\phi_{n+1}),
\end{equation}
where $\tilde{g}(\phi_n,\phi_{n+1})$ is a symmetric quadratic 
polynomial. The form (\ref{first-order-linear}) was used in 
\cite{BOP05} in a search of another family of exceptional 
discretizations which admits {\em translationally invariant 
monotonic kinks} in the discrete $\phi^4$ equation. By the Implicit 
Function Theorem, the first-order difference equation 
(\ref{first-order-difference}) has only one branch of solutions on 
the plane $(\phi_n,\phi_{n+1})$. As a result, one can not construct 
two (decreasing and increasing) sequences $\{ \phi_n 
\}_{n=0}^{\infty}$ in the first-order difference equation 
(\ref{first-order-linear}) for {\em translationally invariant 
single-humped solutions}. Therefore, other exceptional 
discretizations of \cite{BOP05} are irrelevant for localized 
solutions of the discrete NLS equation (\ref{dNLS}).}
\end{Remark}

\section{Existence of traveling solutions near $\omega = (\pi-2)/h^2$ and $c = 1/h$}

We shall consider existence of continuously differentiable solutions 
of the differential advance-delay equation (\ref{advance-delay}) 
with $c \neq 0$. A convenient parametrization of the solution is 
represented by the transformation of variables
\begin{equation}
\phi(z) = \frac{1}{h} \Phi(Z) e^{i \beta Z}, \qquad Z = \frac{z}{h}, 
\end{equation}
and parameters
\begin{equation}
\label{parametrization} \omega = \frac{2}{h} \beta c + \frac{2}{h^2} 
\left( \cos \beta \; \cosh\kappa - 1 \right), \qquad c = \frac{1}{h 
\kappa} \sin \beta \;\sinh \kappa, 
\end{equation}
such that the parameter $h$ is scaled out the new equation for 
$\Phi(Z)$: 
\begin{eqnarray*}
&& 2 i \sin \beta \frac{\sinh \kappa}{\kappa} \Phi'(Z) + 2 \cos 
\beta \cosh \kappa \Phi(Z) \\ && = \Phi(Z+1) e^{i \beta} + \Phi(Z-1) 
e^{-i\beta} + f\left(\Phi(Z-1) e^{-i\beta},\Phi(Z),\Phi(Z+1) e^{i 
\beta}\right), \quad Z \in \mathbb{R}.
\end{eqnarray*}
If the single-humped localized solutions to the differential 
advance-delay equation exist, then the function $\Phi(Z)$ decays in 
$Z$ with the real-valued rate $\kappa$, i.e. $\Phi(Z) \sim 
e^{-\kappa|Z|}$. (Near the bifurcation line $\kappa = 0$, the 
function $\Phi(Z)$ may also contain decaying exponential terms with 
oscillatory tails due to presence of complex eigenvalues but these 
terms decay faster than the tail $\Phi(Z) \sim e^{-\kappa|Z|}$.) The 
boundary on possible existence of traveling solutions corresponds to 
the line $\kappa = 0$, such that the traveling solutions may only 
exist in the exterior domain to the curve
\begin{equation}
\label{boundary-existence} \omega = \frac{2}{h} \beta c -
\frac{4}{h^2} \sin^2 \frac{\beta}{2}, \qquad c = \frac{1}{h} \sin 
\beta,  \qquad \beta \in [0,2\pi].
\end{equation}
The boundary on existence of traveling solutions is shown on Figure 
4 for $c \geq 0$. We apply the technique of the normal form 
reduction from \cite{PR05} when parameters $(\omega,c)$ are close to 
the special values $\omega = (\pi-2)/h^2$ and $c = 1/h$, which 
correspond to $\beta = \pi/2$. By using a modified transformation of 
variables and parameters
$$
\phi(z) = \frac{\epsilon}{h} \Phi(\zeta) \; e^{\frac{i \pi z}{2 h}}, 
\qquad \zeta = \frac{\epsilon z}{h}, \qquad c = \frac{1 + \epsilon^2 
V}{h}, \qquad \omega = \frac{\pi-2 + \epsilon^2 \pi V + \epsilon^3 
\Omega}{h^2},
$$
we rewrite the differential advance-delay equation 
(\ref{advance-delay}) in the equivalent form,
$$
i \left( \Phi(\zeta + \epsilon) - \Phi(\zeta - \epsilon) - 2 
\epsilon \Phi'(\zeta) \right) = \epsilon^3 \left( 2 i V \Phi'(\zeta) 
+ \Omega \Phi(\zeta) \right) - \epsilon^2 f(-i 
\Phi(\zeta-\epsilon),\Phi(\zeta),i \Phi(\zeta+\epsilon)),
$$
where $(\Omega,V) \in \mathbb{R}^2$ are rescaled parameters 
$(\omega,c) \in I_2 \subset \mathbb{R}^2$ and properties P3 and P6 
of the function $f(v,u,w)$ are used. The small parameter $\epsilon > 
0$ defines deviation of parameters $(\omega,c)$ from the values 
$\omega = (\pi-2)/h^2$ and $c = 1/h$ as well as the amplitude and 
localization of the solution $\phi(z)$. By using the formal Taylor 
series expansions of the shift operators $\Phi(\zeta \pm \epsilon)$ 
in powers of $\epsilon$, the differential advance-delay equation for 
$\Phi(\zeta)$ is converted to a third-order ODE which is related to 
the third-order derivative NLS equation \cite{PR05}. The formal 
reduction can be proved with the rigorous technique of the center 
manifold and normal forms (see analysis in \cite{IP06} for kinks in 
the discrete $\phi^4$ equation). 

The formal reduction of the linear and nonlinear parts of the 
differential advance-delay equation for $\Phi(\zeta)$ leads to the 
expansions,
\begin{eqnarray*}
\Phi(\zeta + \epsilon) - \Phi(\zeta - \epsilon) - 2 \epsilon 
\Phi'(\zeta) = \frac{\epsilon^3}{3} \Phi'''(\zeta) + {\rm 
O}(\epsilon^5), 
\end{eqnarray*}
and
\begin{eqnarray*}
f(-i \Phi(\zeta-\epsilon),\Phi(\zeta),i \Phi(\zeta+\epsilon)) =  
\left( \alpha_1 + 2 \alpha_4 - 2 \alpha_5 - 2 \alpha_6 + \alpha_7 
\right) |\Phi|^2 \Phi \\
+ 2 i \epsilon \left( \alpha_2 + 2 \alpha_8 - 2 \alpha_9 \right) 
|\Phi|^2 \Phi'(\zeta) - 2 i \epsilon \left( \alpha_3 - \alpha_8 - 
\alpha_9 + \alpha_{10} \right) \Phi^2 \bar{\Phi}'(\zeta)  + {\rm 
O}(\epsilon^2).
\end{eqnarray*}
By rescaling the amplitude of $\Phi(\zeta)$ one can bring the term 
$|\Phi|^2 \Phi$ to the order of $\epsilon^3$. However, the 
third-order ODE 
$$
\frac{i}{3} \Phi''' - 2 i V \Phi' - \Omega \Phi = |\Phi|^2 \Phi,
$$
has no single-humped localized solutions for any $(\Omega,V) \in 
\mathbb{R}^2$ (see references in \cite{YA03,PR05}). Therefore, the 
necessary condition for existence of single-humped traveling 
solutions is the constraint on parameters of the cubic polynomial 
function (\ref{nonlinearity}):
\begin{equation}
\label{constraint-necessary-1} \alpha_1 + 2 \alpha_4 - 2 \alpha_5 - 
2 \alpha_6 + \alpha_7 = 0.
\end{equation}
When other constraints (\ref{constraints-reductions}) and 
(\ref{constraints-realvalued}) are taken into account, the new 
constraint produces $\alpha_6 = 0$ in the representation 
(\ref{constraints-final}). Under the constraints 
(\ref{constraints-reductions}), (\ref{constraints-realvalued}), and 
(\ref{constraint-necessary-1}) and the normalization 
(\ref{constraint}), no rescaling of the amplitude of $\Phi(\zeta)$ 
is needed and the truncated normal form for traveling solutions 
becomes
\begin{equation}
\label{third-order-ODE} \frac{i}{3} \Phi''' - 2 i V \Phi' - \Omega 
\Phi + 2 i |\Phi|^2 \Phi' + i \gamma \Phi (|\Phi|^2)' = 0,
\end{equation}
where $\gamma = -2 \alpha_3 = 2(\alpha_2 + 2 \alpha_8 - 1)$ is a 
real parameter. Existence of single-humped localized solutions in 
the third-order ODE (\ref{third-order-ODE}) is related to existence 
of embedded solitons in the third-order derivative NLS equation (see 
recent survey in \cite{PY05}). We shall represent the basic facts 
about existence of single-humped localized solutions of the 
third-order ODE (\ref{third-order-ODE}). By using the transformation 
of variables and parameters
$$
\Phi = \lambda \Psi(Z) e^{-ikz}, \qquad Z = \lambda \zeta, \qquad 
\Omega = \frac{2}{3} k \left( \lambda^2 + k^2\right), \qquad V = 
\frac{1}{6} \left( \lambda^2 - 3 k^2 \right),
$$
we rewrite the ODE (\ref{third-order-ODE}) in the form,
\begin{eqnarray}
\label{ODE} \frac{i \lambda}{3} \left(\Psi''' - \Psi' + 6 |\Psi|^2 
\Psi' + 3 \gamma \Psi (|\Psi|^2)' \right) + k \left(\Psi'' - \Psi + 
2 |\Psi|^2 \Psi \right) = 0,
\end{eqnarray}
where the real-valued parameters $(\lambda,k)$ are arbitrary. The 
value $\lambda = 0$ defines the bifurcation cutoff in the family of 
localized solutions of the ODE (\ref{ODE}), which is equivalent to 
the curve $V = V_{\rm thr}(\Omega)$ in the parameter plane 
$(\Omega,V)$, where
\begin{equation}
\label{threshold} V_{\rm thr}(\Omega) = - 
\frac{(3\Omega)^{2/3}}{2^{5/3}}.
\end{equation}
Localized solutions may only exist above the threshold $V > V_{\rm 
thr}(\Omega)$ for any $\Omega \in \mathbb{R}$. The asymptotic 
approximation of the threshold (\ref{threshold}) is shown on Figure 
4 by dotted curve. The threshold (\ref{threshold}) is an asymptotic 
approximation of the exact curve (\ref{boundary-existence}). 

\begin{itemize}
\item When $\gamma = 0$, the third-order ODE (\ref{ODE}) is related to the 
integrable Hirota equation \cite{H73}, which admits the exact 
single-humped localized solution $\Psi = {\rm sech}(Z)$. The 
single-humped solution exists everywhere on the two-parameter plane 
$(\Omega,V)$ above the threshold (\ref{threshold}). 

\item When $\gamma = 1$, the third-order ODE (\ref{ODE}) is related to the 
integrable Sasa-Satsuma equation \cite{SS91}, which also admits the 
exact localized solutions everywhere on the two-parameter plane 
$(\Omega,V)$ above the threshold (\ref{threshold}). (The exact 
solution and its properties are given in Sections 4-5 of 
\cite{SS91}.) The localized solution has a single-humped profile for 
$V_{\rm thr}(\Omega) < V < V_{\rm humps}(\Omega)$ and a 
double-humped profile for $V > V_{\rm humps}(\Omega)$ and $\Omega 
\neq 0$, where 
\begin{equation}
\label{humps} V_{\rm humps}(\Omega) = - \frac{(3\Omega)^{2/3}}{3 
2^{4/3}}.
\end{equation}
The threshold (\ref{humps}) in the ODE (\ref{ODE}) corresponds to 
the condition $\lambda^2 = k^2$. It is shown on Figure 4 by 
dashed-dotted curve. When $\Omega \to 0$ and $V > 0$, the distance 
between two humps diverge and the ODE (\ref{ODE}) has again the 
single-humped solution for $\Omega = 0$ and $V > 0$ with $\Psi = 
\frac{1}{\sqrt{2}} {\rm sech}(Z)$. This solution corresponds to $k = 
0$ in the ODE (\ref{ODE}).

\item When $\gamma \in \mathbb{R}$, the third-order ODE (\ref{third-order-ODE}) has 
the exact single-humped localized solution for $\Omega = 0$ and $V > 
0$ with $\Psi = \frac{1}{\sqrt{1 + \gamma}} {\rm sech}(Z)$. The 
exact solution matches the member of the family of exact solutions 
in the two integrable cases $\gamma = 0$ and $\gamma = 1$. It is 
shown in \cite{PY05} by numerical analysis of the kernel of the 
linearization operator (see Sections 3-4 in \cite{PY05}) that the 
family of single-humped localized solutions for $\gamma \in 
\mathbb{R} \backslash \{ 0,1\}$ is isolated from other families of 
localized solutions, i.e. the one-parameter family with $\Omega = 0$ 
and $V > 0$ can not be continued in the two-parameter plane 
$(\Omega,V)$ as a single-humped localized solution.
\end{itemize}

\begin{figure}[htbp]
\begin{center}
\includegraphics[height=7cm]{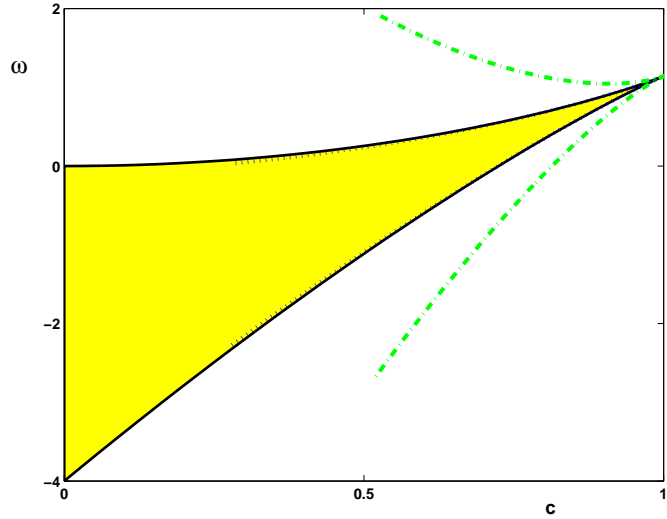} 
\end{center}
\caption{The boundary (\ref{boundary-existence}) of the domain of 
existence for traveling solutions of the differential advance-delay 
equation (\ref{advance-delay}) and asymptotic approximations for the 
threshold on existence (\ref{threshold}) (dotted curve) and the 
threshold on two-humped solutions (\ref{humps}) (dashed-dotted 
curve).}
\end{figure}

Results of the third-order ODE (\ref{third-order-ODE}) give only 
necessary condition on existence of traveling single-humped solution 
in the differential advance-delay equation (\ref{advance-delay}) 
near the special values $\omega = (\pi-2)/h^2$ and $c = 1/h$, i.e. 
if the smooth solution $\phi(z)$ to the differential advance-delay 
equation (\ref{advance-delay}) exists, then it matches the 
analytical solutions of the third-order ODE (\ref{third-order-ODE}). 
Persistence proof of true localized solutions is a delicate problem 
of rigorous analysis, which is left open even for traveling kinks of 
the discrete $\phi^4$ equation \cite{IP06}. (For comparison, 
persistence of solutions with exponentially small non-localized 
oscillatory tails was rigorously proved near the normal form 
equation in \cite{IP06}.) 

When the discrete NLS equation (\ref{dNLS}) is the integrable AL 
lattice (\ref{AL}), the exact traveling solutions exist in the form 
$\Phi(Z) = \sinh \kappa \; {\rm sech}(\kappa Z)$ for any $\kappa > 
0$ and $\beta \in [0,2 \pi]$. These solutions match the exact 
solutions for the Hirota equation for $V > V_{\rm thr}(\Omega)$ and 
$\Omega \in \mathbb{R}$ in the form $\Psi(\zeta) = {\rm 
sech}(\zeta)$ with the correspondence $\kappa = \epsilon \lambda$. 
Additionally, when the discrete NLS equation (\ref{dNLS}) has the 
nonlinear function 
$$
f = (1 - \alpha) |u_n|^2 (u_{n+1} + u_{n-1}) + \alpha u_n^2 
(\bar{u}_{n+1} + \bar{u}_{n-1}), \qquad \alpha \neq 0,
$$ 
the exact traveling solution exists for $\kappa > 0$ and $\beta = 
\frac{\pi}{2}$ in the form 
$$
\Phi(Z) = \frac{1}{\sqrt{1 - 2 \alpha}} \sinh \kappa \; {\rm 
sech}(\kappa Z).
$$
This solution matches the exact solution of the third-order ODE 
(\ref{third-order-ODE}) for $\Omega = 0$ and $V > 0$ in the form 
$\Psi(\zeta) = \frac{1}{\sqrt{1 + \gamma}} {\rm sech}(Z)$ (where 
$\gamma = - 2\alpha_3$ and $\alpha = \alpha_3$). We do not know if 
there are any other examples of the discrete NLS equation 
(\ref{dNLS}) with the nonlinear function (\ref{nonlinearity}) which 
admit families of single-humped traveling solutions which match the 
analytical solutions of the third-order ODE (\ref{third-order-ODE}). 
See \cite{Baldwin} for symbolic tests of explicit $\tanh$ and ${\rm
sech}$ solutions of differential advance-delay equations.

\section{Conclusion}

We have considered stationary and traveling solutions of the 
discrete NLS equation (\ref{dNLS}) with the ten-parameter cubic 
nonlinearity (\ref{nonlinearity}) under the normalization constraint 
(\ref{constraint}). We have proved in Lemma \ref{lemma-1} that four 
constraints (\ref{constraints-reductions}) are sufficient for 
reduction of the second-order difference equation 
(\ref{second-order-difference}) to the first-order difference 
equation (\ref{first-order-difference}). The main result 
(Proposition \ref{theorem-existence}) says that this reduction gives 
a sufficient condition for existence of translationally invariant 
stationary solutions that converge to the stationary solutions 
(\ref{soliton}) with $c = 0$ in the continuum limit $h \to 0$. 

Furthermore, we have proved in Lemma \ref{lemma-10} that two 
additional constraints (\ref{constraints-realvalued}) are sufficient 
for uniqueness of real-valued stationary localized solutions in the 
first-order difference equation (\ref{first-order-difference}). We 
have finally found in Section 4 that one more constraint gives the 
necessary condition for existence of traveling solutions in the 
differential advance-delay equation (\ref{advance-delay}) near the 
special values $\omega = (\pi-2)/h^2$ and $c = 1/h$. Combining all 
constraints (\ref{constraints-reductions}), 
(\ref{constraints-realvalued}), and (\ref{constraint-necessary-1}) 
and the normalization constraint (\ref{constraint}), we have a 
two-parameter family of the discrete NLS equation with the nonlinear 
function
\begin{equation}
\label{translation-invariance} f = (1 - \alpha - 2\beta) |u_n|^2 
(u_{n+1} + u_{n-1}) + \alpha u_n^2 (\bar{u}_{n+1} + \bar{u}_{n-1}) + 
\beta (|u_{n+1}|^2 + |u_{n-1}|^2) (u_{n+1} + u_{n-1}),
\end{equation}
where $(\alpha,\beta)$ are two real-valued parameters. When $\alpha 
= \beta = 0$, the model (\ref{translation-invariance}) is the AL 
lattice (\ref{AL}) with the two-parameter family of traveling 
solutions in the domain (\ref{boundary-existence}). When $\alpha = 
0$, the model (\ref{translation-invariance}) is related to the 
integrable Hirota equation (\ref{third-order-ODE}) with $\gamma = 
0$. When $\alpha = -\frac{1}{2}$, the model 
(\ref{translation-invariance}) is related to the integrable 
Sasa-Satsuma equation (\ref{third-order-ODE}) with $\gamma = 1$. 
When $\beta = 0$, the model (\ref{translation-invariance}) admits 
one-parameter family of exact traveling solutions on the line 
$\omega = (\pi ch-2)/h^2$ and $c > 1/h$. 

Although the discrete NLS equation (\ref{dNLS}) with the nonlinear 
function (\ref{translation-invariance}) has translationally 
invariant stationary solutions (\ref{stationary}), the existence of 
traveling solutions (\ref{travelling}) for any $c \neq 0$ is an open 
problem, which is left beyond the scopes of the present manuscript. 
Persistence of traveling solutions for small values of $c$ can not 
be proved since infinitely many resonances with infinitely many 
Stokes constants appear in the limit $c \to 0$ \cite{OPB05}. (It was 
shown in \cite{OPB05} with numerical computations of the leading 
Stokes constant that none of three particular discrete $\phi^4$ 
lattices exhibit families of traveling solutions bifurcating from 
the family of translationally invariant stationary solutions.) 
Although existence of translationally invariant stationary solutions 
gives only the necessary condition for persistence of traveling 
solutions, the integrable AL lattice represents at least one example 
of the discrete NLS equation (\ref{AL}) where persistence of 
traveling solutions can in principle occur. 

Similarly, persistence of traveling solutions near the integrable 
normal form (\ref{third-order-ODE}) can not be proved since 
oscillatory tails are generic near the values $\omega = (\pi-2)/h^2$ 
and $c = 1/h$ \cite{IP06}. While the AL lattice 
(\ref{translation-invariance}) with $\alpha = \beta = 0$ has exact 
traveling solutions between the two limiting cases $c = 0$ and $c = 
1/h$, it is not clear if the translationally invariant NLS lattice 
(\ref{translation-invariance}) with $\alpha,\beta \neq 0$ exhibits 
any families of traveling solutions on the plane $(\omega,c)$. If 
such solutions exist for sufficiently small values of $h$, these 
solutions converge to the family (\ref{soliton}) in the continuum 
limit $h \to 0$. This problem can be considered by means of 
numerical solutions of the differential advance-delay equation 
(\ref{advance-delay}) similarly to the numerical works 
\cite{Hump,Champ}.

{\bf Acknowledgement.} This work was initiated by discussions with 
I. Barashenkov, P. Kevrekidis and A. Tovbis. The work was partly 
supported by the France--Canada SSHN Advanced Level Fellowship and 
by the PREA grant.


\begin{thebibliography}{99}
\bibitem[AEHV05]{Hump} K.A. Abell, C.E. Elmer, A.R. Humphries, and
E.S.V. Vleck, "Computation of mixed type functional differential
boundary value problems", SIAM J. Applied Dynamical Systems {\bf 4},
755--781 (2005)

\bibitem[AM03]{AM03} M.J. Ablowitz and Z.H. Musslimani, "Disrcete spatial solitons
in a diffraction--managed nonlinear waveguide array: a unified
approach", Physica D {\bf 184}, 276--303 (2003)

\bibitem[BOP05]{BOP05} I.V. Barashenkov, O.F. Oxtoby, and D.E.
Pelinovsky, "Translationally invariant discrete kinks from
one-dimensional maps", Physical Review E {\bf 72}, 035602(R) (2005)

\bibitem[BGH04]{Baldwin} D. Baldwin, U. Goktas, and W. Hereman, 
"Symbolic computation of hyperbolic tangent solutions for nonlinear 
differential-difference equations", Computer Physics Communications
{\bf 162}, 203-217 (2004). 

\bibitem[BGKM91]{MacKay} C. Baesens, J. Guckenheimer, S. Kim, and R.S. MacKay, "Three 
coupled oscillators: Mode-locking, global bifurcations and toroidal 
chaos", Physica D {\bf 49}, 387--475 (1991)

\bibitem[C06]{Champ} A. Champneys, personal communication (2006). 

\bibitem[CKKS93]{CKKS93} C. Claude, Y.S. Kivshar, O. Kluth, and
K.H. Spatschek, "Moving localized modes in nonlinear lattices", 
Phys. Rev. B {\bf 47}, 14228--14232 (1993).

\bibitem[DKY05]{DKY05} S.V. Dmitriev, P.G. Kevrekidis, and
N. Yoshikawa, "Discrete Klein--Gordon models with static kinks free 
of the Peierls--Nabarro potential", J. Phys. A.: Math. Gen. {\bf 
38}, 7617--7627 (2005).

\bibitem[FP99]{FP99} G. Friesecke and R.L. Pego, "Solitary waves on FPU lattices: I. Qualitative 
properties, renormalization and continuum limit", Nonlinearity {\bf 
12}, 1601--1627 (1999)

\bibitem[H73]{H73} R. Hirota, "Exact envelope-soliton solutions of a nonlinear
wave equation", J. Math. Phys. {\bf 14}, 805--809 (1973).

\bibitem[IP06]{IP06} G. Iooss and D. Pelinovsky, 
"Normal form for travelling kinks in discrete Klein--Gordon 
lattices", Physica D, accepted for publication (2006)

\bibitem[K03]{K03} P.G. Kevrekidis, ``On a class of discretizations
of Hamiltonian nonlinear partial differential equations'', Physica D
{\bf 183}, 68--86 (2003).

\bibitem[KRB01]{Kevrekidis-review} P.G. Kevrekidis, K.O. Rasmussen, and A.R. Bishop,
``The discrete nonlinear Schr\"{o}dinger equation: a survey of
recent results'', Int. J. Mod. Phys. {\bf 15}, 2833--2900 (2001).

\bibitem[OPB05]{OPB05} O.F. Oxtoby, D.E. Pelinovsky, and I.V. Barashenkov, 
"Travelling kinks in discrete $\phi^4$ models", Nonlinearity {\bf 
19}, 217--235 (2006)

\bibitem[OWW04]{symplecticNLS} M. Oliver, M. West, and C. Wulff, "Approximate momentum
conservation for spatial semidiscretizations of semilinear wave
equations", Numer. Math. {\bf 97}, 493--535 (2004)

\bibitem[OJE03]{OJE03} M. Oster, M. Johansson, and A. Eriksson, "Enhanced mobility of strongly
localized modes in waveguide arrays by inversion of stability",
Phys. Rev. E {\bf 67}, 056606 (2003)

\bibitem[P05]{pankov} A. Pankov, {\em Travelling Waves and Periodic Oscillations in Fermi--Pasta--Ulam Lattices}
(Imperial College Press, London, 2005)

\bibitem[PR05]{PR05} D.E. Pelinovsky and V.M. Rothos, "Bifurcations
of traveling wave solutions in the discrete NLS equations", Physica
D {\bf 202}, 16--36 (2005).

\bibitem[PY05]{PY05} D. Pelinovsky and J. Yang, "Stability analysis of embedded 
solitons in the generalized third-order NLS equation", Chaos {\bf 
15}, 037115 (2005).

\bibitem[SS91]{SS91} N. Sasa and J. Satsuma,
``New type of soliton solutions for a higher-order nonlinear 
Schr\"odinger equation.'' J. Phys. Soc. Japan 60, 409-417 (1991).

\bibitem[T00]{Tovbis} A. Tovbis, "On approximation of stable and unstable manifolds and the Stokes phenomenon", 
Contemp. Math. {\bf 255}, 199--228 (2000)

\bibitem[YA03]{YA03} J. Yang and T.R. Akylas, "Continuous families
of embedded solitons in the third-order nonlinear Schr\"{o}dinger 
equation", Stud. Appl. Math. {\bf 111}, 359--375 (2003).

\end{thebibliography}
\end{document}